\documentclass[twocolumn]{aastex63}
\usepackage{graphicx}
\usepackage{subfigure}
\usepackage{amsmath}    
\usepackage{amssymb}    
\usepackage{bm}
\usepackage{url}
\usepackage{upgreek}
\usepackage{xspace}

\newcommand{\deld}{\ensuremath{\delta d}\xspace}
\newcommand{\ppcc}{$\,$pc$\,$cm$^{-3}$\xspace} 
\newcommand{\tsys}{T$_{\rm sys}$\xspace}       
\newcommand{\bandiii}{band~3\xspace}
\newcommand{\bandiv}{band~4\xspace}
\newcommand{\bandv}{band~5\xspace}
\newcommand*{\logten}{\mathop{\log_{10}}}
\newcommand{\peakf}{\ensuremath{\nu_p}\xspace}
\newcommand{\cnot}{\ensuremath{C_0}\xspace}
\newcommand{\wten}{\ensuremath{W_{10}}\xspace}
\def\magn{XTE~J1810$-$197\xspace}
\def\mpsr{PSR~J1809$-$1943\xspace}

\accepted{\today}

\shorttitle{Magnetar XTE~J1810$-$197: Spectro-temporal evolution of average radio emission}
\shortauthors{Maan et al.}

\begin{document}

\title{Magnetar XTE~J1810$-$197: Spectro-temporal Evolution of Average Radio Emission}

\correspondingauthor{Yogesh Maan}
\email{ymaan@ncra.tifr.res.in}
\author[0000-0002-0862-6062]{Yogesh Maan}
\affil{National Centre for Radio Astrophysics, Tata Institute of Fundamental Research, Post Bag 3, Ganeshkhind, Pune - 411007, India}
\affil{ASTRON, the Netherlands Institute for Radio Astronomy, Postbus 2, 7990 AA, Dwingeloo, The Netherlands}
\author{Mayuresh P. Surnis}
\affiliation{Jodrell Bank Centre for Astrophysics, Department of Physics and Astronomy, The University of Manchester, Manchester, M139PL, UK}
\author{Bhal Chandra Joshi}
\affil{National Centre for Radio Astrophysics, Tata Institute of Fundamental Research, Post Bag 3, Ganeshkhind, Pune - 411007, India}
\author{Manjari Bagchi}
\affiliation{The Institute of Mathematical Sciences, C.I.T. Campus, Taramani, Chennai, 600113, India}
\begin{abstract}
We present the long-term spectro-temporal evolution of the average radio
emission properties of the magnetar \magn (\mpsr) following its
most recent outburst in late 2018. We report the results from two and a half
years of monitoring campaign with the upgraded Giant Metrewave Radio Telescope
carried out over the frequency range of 300$-$1450 MHz. Our observations show
intriguing time variability in the average profile width,
flux density, spectral index
and the broadband spectral shape. While the average profile width appears to
gradually decrease at later epochs, the flux density shows multiple episodes
of radio re-brightening over the course of our monitoring. Our systematic
monitoring observations reveal that the radio spectrum
has steepened over time, resulting in evolution from a magnetar-like to
a more pulsar-like spectrum. A more detailed analysis reveals that the radio
spectrum has a turnover, and this turnover shifts towards lower frequencies
with time. We present the details of our analysis leading to these results, and
discuss our findings in the context of magnetar radio emission mechanisms as well
as potential manifestations of the intervening medium. We also briefly discuss
whether an evolving spectral turnover could be an ubiquitous property of
radio magnetars.
\end{abstract}
\keywords{Stars: magnetars, pulsars: general, pulsars: individual (J1809$-$1943),
radiation mechanisms: non-thermal, ISM: general}

\section{Introduction} \label{sec-intro}
Magnetars are highly magnetized ($10^{13} - 10^{15}$ G) neutron stars having long spin periods (1.3 $-$ 12 s). They are powered by the decay of their internal magnetic field rather than the rotational kinetic energy \citep{DT92} and some of them show radio emission after a high-energy burst or flare \citep[e.g.][]{Camilo06}. Currently, five magnetars are known to exhibit transient periodic radio emission, and another one produces isolated radio bursts \citep[see McGill magnetar catalog\footnote{\url{http://www.physics.mcgill.ca/~pulsar/magnetar/main.html}};][]{ok14}. They typically display time variability in many characteristics like radio flux density, spectra and integrated profile shapes. The radio emission typically has a flat spectrum. This has enabled the study of the radio emission at radio frequencies as high as 353\,GHz for some of the radio-loud magnetars \citep{Torne22}. The very high frequency observations indicate that the radio spectrum has a high frequency turn-up. However, the low frequency end of the radio spectrum has not been explored through regular monitoring observations in the past. 

\magn was discovered during an outburst in 2004 \citep{Ibrahim04} and was the first radio-loud magnetar \citep{Camilo06}. Following the onset, the radio pulsations lasted for nearly three years before becoming undetectable around 2008 \citep{Camilo16}. During the radio-loud episode, the magnetar showed time variable flux density, pulse profile and spectral index \citep{Camilo07a,Camilo07b,Lazaridis08}. The radio flux density declined initially. Then it seemed to vary around a steady value before the radio pulsations stopped abruptly \citep{Camilo16}. 

The current activity began in late 2018 with an intense episode in radio and X-ray \citep{Lyne18,Gotthelf18}. Radio pulsations were soon detected at a very broad range of radio frequencies and follow-up observations have been continuing since then \citep{Joshi18,Trushkin19,Dai19,Torne20}. Similar to the last outburst, the magnetar has shown variations in profile shape, flux density as well as single-pulse properties \citep{Dai19,Levin19,Maan19b,Caleb21}. In our previous paper \citep{Maan19b}, we primarily reported on the single-pulse properties, along with the low-frequency radio flux density and spectrum of the magnetar close to the outburst. As mentioned earlier, a broad-band radio spectrum is crucial in order to understand the emission mechanism. Given most of the current monitoring campaigns observe the magnetar at frequencies higher than 1 GHz, our campaign has made use of the upgraded Giant Metrewave Radio Telescope \citep[GMRT;][]{Gupta17} to cover a frequency range of 300$-$1450 MHz. 

In this paper, we report on the temporal evolution of the pulsed flux density, spectral index as well as the average profile width from our monitoring campaign. We describe the observations in Section \ref{sec-obs}, results and analysis in Sections \ref{sec-flux} and \ref{sec-turnover}, and discuss the implications in Section \ref{sec-disc}

\begin{figure*}
\centering
\includegraphics[width=0.8\textwidth]{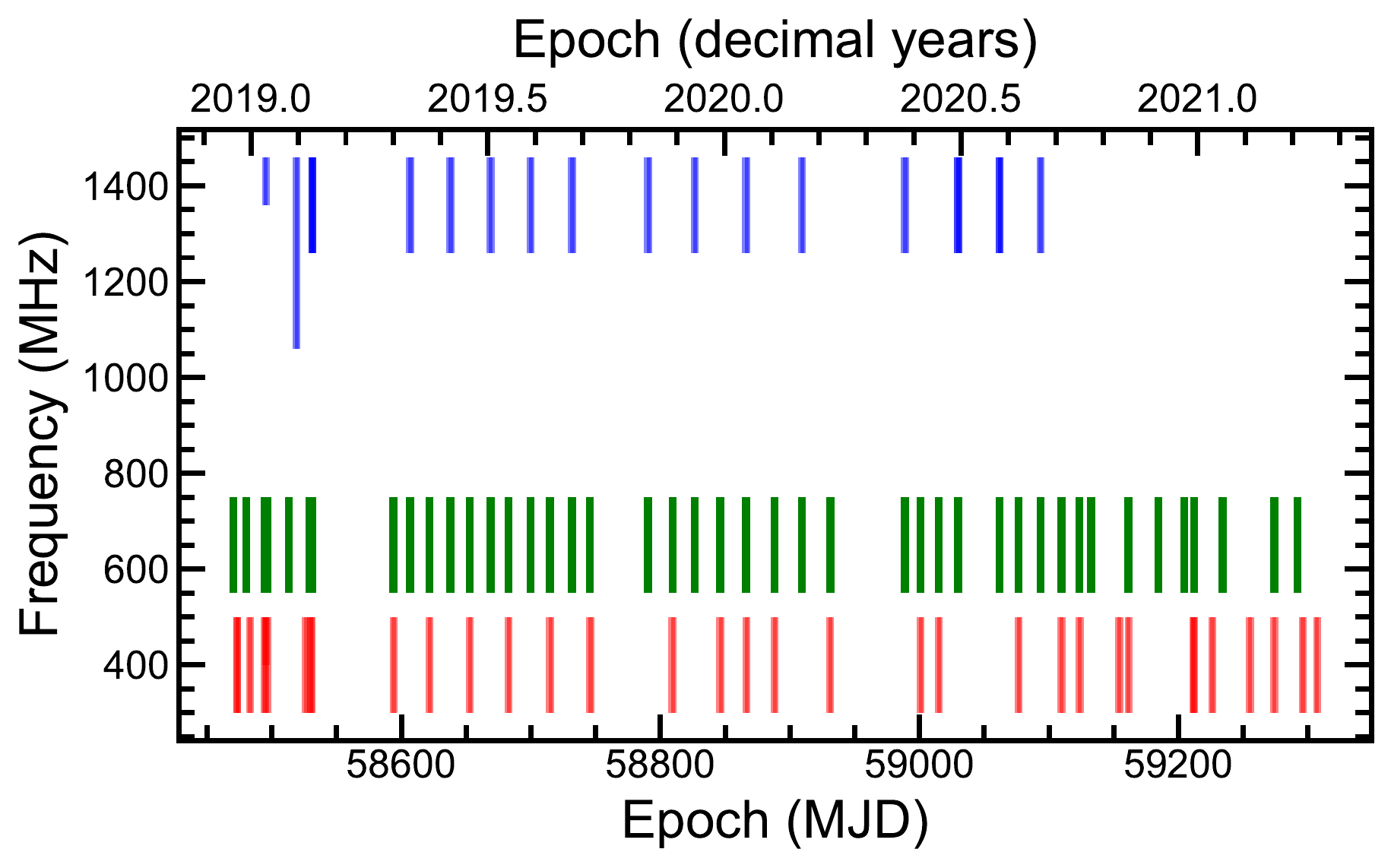}
\caption{A summary of the observations presented in this paper. The individual
vertical bars indicate the frequency range covered by an observation at a
particular epoch. The red, green and blue colored bars represent bands 3
(300$-$500\,MHz), 4 (550$-$750\,MHz) and 5 (typically 1260$-$1460\,MHz) of GMRT, and
the same color convention has been followed in the rest of the paper.
The band~5 observations at the first two epochs utilized bandwidths of
100 and 400\,MHz, respectively, and 200\,MHz at all the other epochs.}
\label{fig-obs}
\end{figure*}
\section{Observations and data reduction} \label{sec-obs}
After successful detection and follow-up of the magnetar \magn at a
number of epochs in late 2018 and early 2019 \citep{Joshi18,Maan19b}
using the Director's Discretionary Time allocations, we started a
low-frequency monitoring campaign on this source with the GMRT.
The monitoring observations utilized bands~3, 4 and 5 of the GMRT with
typical center frequencies of 400, 650 and 1360\,MHz, respectively, and
200\,MHz bandwidths at each of the bands.
Depending on the spectral and
temporal evolution of the
magnetar's flux density, the observations were conducted in a number of
configurations: simultaneous observations in two or three bands at a few
epochs by combining GMRT antennae in two or three sub-arrays,
near-simultaneous observations in two of the
bands by observing in one band after the other, or just single band
observations. The near-simultaneous observations always included band~4
as one of the two bands. After September 2020, the monitoring observations
were limited to bands~3 and 4.
Figure~\ref{fig-obs} shows the cadence and a summary of all the observations
presented in this paper.
\par
The observation on MJD~58495 was conducted simultaneously in bands~3,
4 and 5, and the bandwidth in each of the band was limited to 100\,MHz in
this configuration. However, only band~5 data could be used from this
observation due to severe contamination by radio frequency interference (RFI)
in the other two bands. As apparent in Figure~\ref{fig-obs}, the band~5
observation on MJD~58519 employed 400\,MHz bandwidth. A bandwidth of 200\,MHz
was utilized in all the subsequent band~5 observations, many of which were
conducted in an observing mode that facilitated simultaneous dual-band observations
and limited the bandwidth to 200\,MHz.
Initially, the observing setup constituted
8192 channels and 0.328 or 0.655\,ms time resolution in band~3, and 4096 channels
and 0.164\,ms sampling time in bands~4 and 5. Since November 2020, we used
an observing mode wherein the data are coherently dedispersed at a dispersion
measure (DM) of 178.85\,\ppcc in real time and then recorded to the disk with
1024 frequency sub-bands, and sampling times of 0.164 and 0.081\,ms in bands~3
and 4, respectively.
\par
The recorded data for all the bands are processed through a series of data
reduction steps. For the observations prior to November 2020, we use
SIGPROC's \texttt{dedisperse} to sub-band the data to 1024 channels wherein
the data within each sub-band is dedispersed using a DM of 178.85\,\ppcc.
As mentioned above, observations after November 2020 were already recorded
with 1024 coherently dedispersed sub-bands. The data from the individual
epochs are then subjected
to size reduction by down-sampling from 16\,bits to 8\,bits per sample using
\texttt{digifil}, and RFI excision using
\texttt{RFIClean}\footnote{\url{https://github.com/ymaan4/RFIClean}}
\citep{MvLV21} and
\texttt{rfifind} from the pulsar search and analysis toolkit \textsl{PRESTO}
\citep{Ransom02}. The resultant data are folded using \texttt{prepfold}
from \textsl{PRESTO} and the timing parameters from \citet{Levin19}, with
1024 bins, 128 sub-bands and typically 64 sub-intervals. This \textsl{PRESTO}
utility outputs partially folded data
along with the several pieces of information, including the period and DM
values which maximize the average profile signal-to-noise ratio (S/N),
in files with extensions \texttt{pfd}, which we refer to as pfd-files here onwards.
\section{Analysis and Results: Flux density, spectral index, profile width and their temporal evolution} \label{sec-flux}
\subsection{Calibration procedure}
In a number of our observing sessions, we had carried out scans on flux
calibrators (3C286 and 3C48) as well as a few degrees away from them. Using
these on-source and off-source calibrator observations, we estimated the system
equivalent flux density (hereafter SEFD, defined as the ratio of the system
temperature and the gain, \tsys/G) per GMRT dish as a function of frequency and
fitted a polynomial to these measurements. For \bandiv,
the estimated SEFD compares well with the polynomial
fits for the sensitivity used in the GMRT exposure time calculator
(ETC\footnote{\url{http://www.ncra.tifr.res.in:8081/~secr-ops/etc/etc.html}}).
For \bandv, only a few calibrator scans were available and some of
these were heavily contaminated by RFI. Nevertheless, we could estimate the
SEFD for \bandv using one set of calibrator scans, and it was found to be
slightly offset from the polynomial fit used in ETC.
For \bandiii, the SEFD
could be estimated using calibrator scans at several epochs, however, the
SEFD measurements show a large spread around the mean value.
\begin{figure*}
\centering
\includegraphics[width=0.8\textwidth]{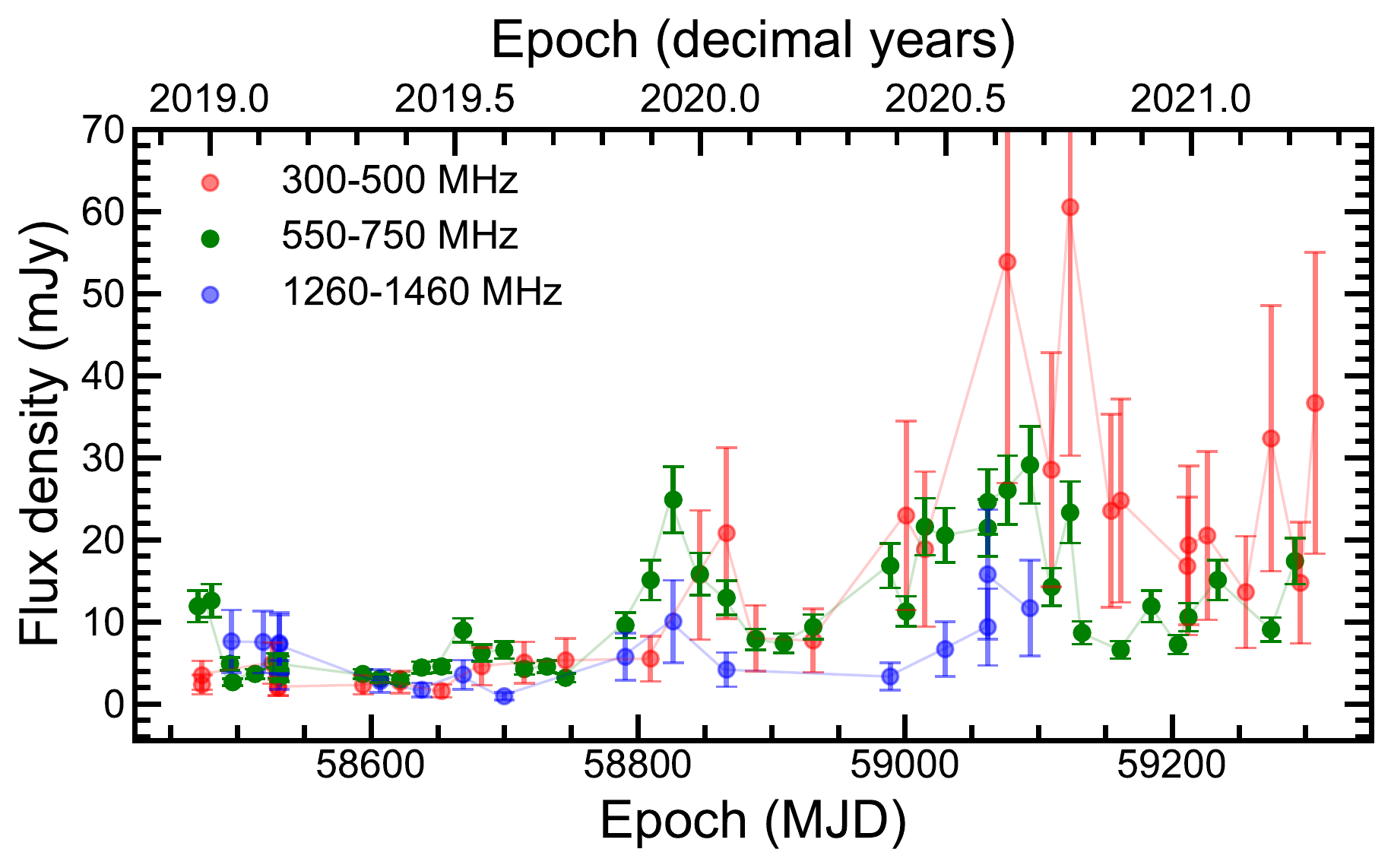}
\caption{Period averaged flux densities as a function of epoch,
measured in the individual bands 3 (300$-$500\,MHz), 4 (550$-$750\,MHz)
and 5 (1260$-$1460\,MHz).}
\label{fig-flux}
\end{figure*}
\par
We read the partially folded data from the pfd-files in python using
the class \texttt{pfd} provided in \textsl{PRESTO}, and re-fold
these using the best period and DM values suggested by \texttt{prepfold}.
Despite the RFI excision using RFIClean as well as \texttt{rfifind}, some faint
RFI becomes visible only in the partially folded data, which primarily
appears in the form of baseline variations. It is important to get rid
of or correct for these baseline variations to correctly estimate the
flux density using the radiometer equation \citep{Handbook04}.
However, the long period
of the magnetar also does not help in averaging out these variations.
To remedy the situation, the off-pulse
regions of the profiles from the sub-intervals and sub-bands are used
to estimate robust statistics and then subjected to a threshold-based
identification of outliers. Additionally, 12.5\% of the sub-bands on
either side of the band, as well as the sub-bands in the frequency
range 355$-$385\,MHz (which is often contaminated by the signals from the MUOS satellites),
are also considered as outliers. For severely RFI contaminated data,
the sub-interval profiles are also inspected by eye to identify the ones
with visibly contaminated baselines. The outlier sub-intervals and
sub-bands thus identified are excluded from any further processing.
These data are then averaged fully over time and to a pre-specified
number of final sub-bands, which is typically 4 for \bandiv,
4 or 2 for \bandiii and just 1 (i.e., averaged over the full bandwidth) for
\bandv.
The off-pulse regions of the sub-banded profiles were also fitted
by a 3$^{rd}$ order (9$^{th}$ order for profiles with S/N$>$500)
polynomial to get rid of any remaining baseline
variations. These sub-banded and baseline-corrected average profiles
are then flux calibrated
by estimating the mean and standard deviation in the off-pulse
region and using the radiometer equation with the SEFD estimates
described above. The observing time and bandwidth is appropriately
accounted for the sub-intervals and sub-bands/channels that are
identified as outliers by \texttt{rfifind} as well as in the above
post-processing. The frequency-dependent sky temperature towards the
source is estimated using the
\texttt{skytempy}\footnote{\url{https://libraries.io/pypi/skytempy}}
package, which is based on the reprocessing of the \citet{Haslam82}
408\,MHz map by \citet{Remazeilles15}. Assuming receiver temperatures
of 85, 87 and 62\,K at the centers of the bands 3, 4 and 5, respectively,
a constant gain throughout the bands and the sky temperature estimates
from above, the SEFD is re-estimated towards the source before using
in the radiometer equation. Furthermore, we assume that the signals
from different antennae in a sub-array are added fully coherently,
i.e., the gain of a sub-array is linearly proportional to the number
of antennae.
\begin{figure*}
\centering
\includegraphics[width=0.8\textwidth]{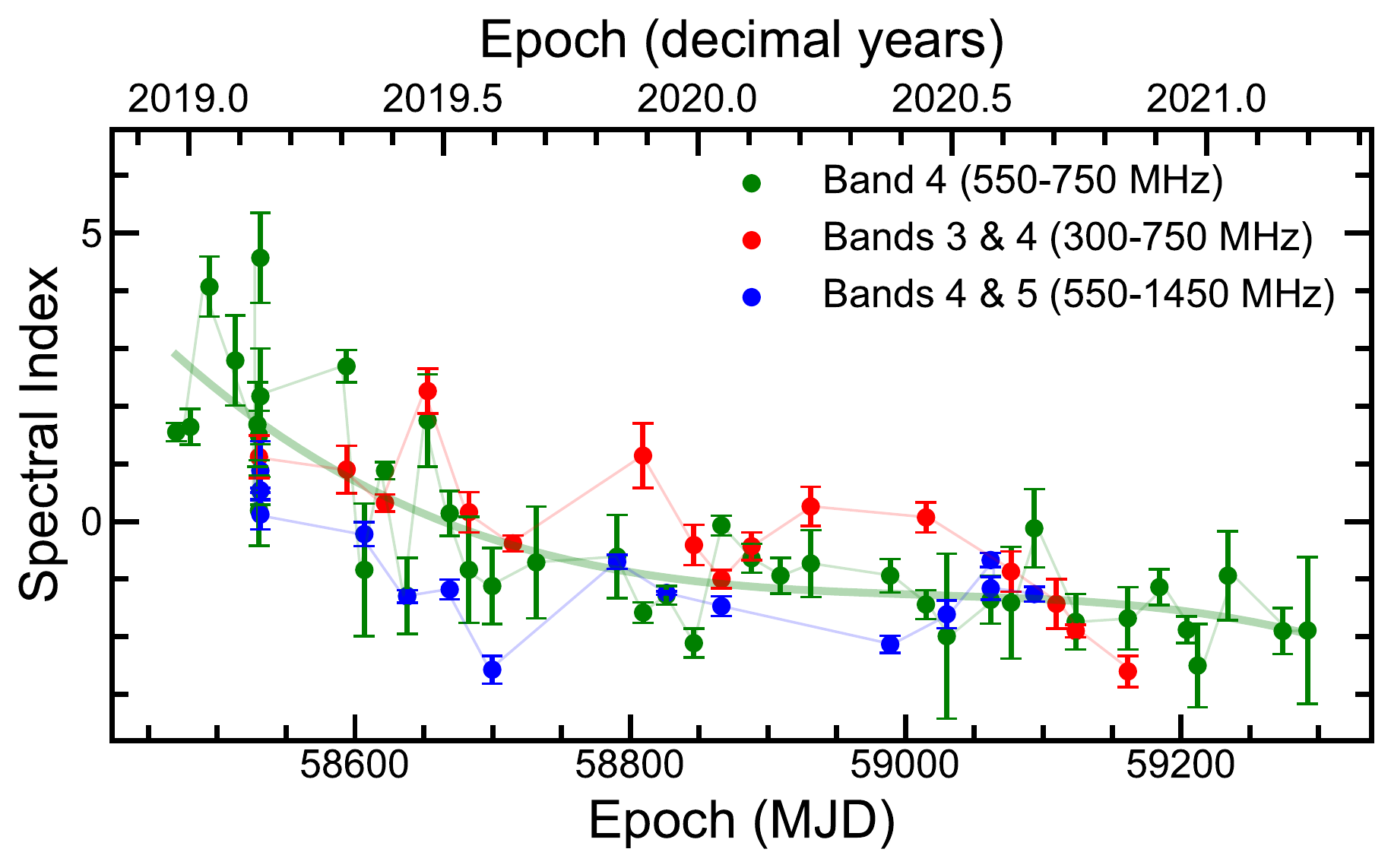}
\caption{Power-law spectral indices estimated using the period averaged
flux densities measured in band~4, bands 3 \& 4, and bands 4 \& 5.
The light-green colored thick, continuous curve shows a
$3^{rd}$-order polynomial fit to the band~4 estimates, only to demonstrate
the temporal evolution of the spectral indices and guide the eye.}
\label{fig-alpha}
\end{figure*}
\subsection{Flux density}
By integrating the area under these flux-calibrated, sub-banded profiles,
we estimate the period-averaged flux densities for each of the sub-bands.
The corresponding uncertainties are estimated by assuming 5\% and 10\%
uncertainties in SEFD and the system temperature, respectively. 
The period-averaged flux density over the entire band is then estimated
by averaging the corresponding sub-band estimates. The flux density of the magnetar
measured in the three bands is shown as a function of epoch in
Figure~\ref{fig-flux}.
\par
We note here that there are primarily three factors which could have affected
our flux density measurements. First, as mentioned earlier, we have assumed
that the signals from different antennae in a sub-array are added fully
coherently. In practice, ionospheric or weather conditions might introduce
disturbances in an otherwise coherently phased array. Specifically at low
radio frequencies, a full coherence might also not be always achievable. Any such
deviation from a fully coherent addition would have resulted in an additional
systematic uncertainty.
\par
Second, as \magn is a long period pulsar, any low-level baseline variations
in the off-pulse region might result in slightly over-estimated standard
deviation measurements, and hence, under-estimated flux densities. From
all the scrutiny and care taken in excising RFI and baseline correction
mentioned in the previous sub-section as well as visual inspection of
profiles, we expect this issue to have affected our measurements only at
a few epochs.
\par
Third, the large spread in the SEFD measurement at \bandiii implies a
large uncertainty in our \bandiii flux density measurements. The \bandv
SEFD was estimated using calibrator scans at only one epoch and it was
found to be offset from that used in ETC by about 20\%. For these reasons,
we have assumed the uncertainties of our \bandiii and \bandv flux density
measurements to be 50\%. We note that our \bandv flux density measurements
are largely consistent with those measured using the Jodrell bank telescope
at similar epochs \citep{Caleb21}.
\begin{figure*}
\centering
\subfigure[]{\includegraphics[width=0.3\textwidth]{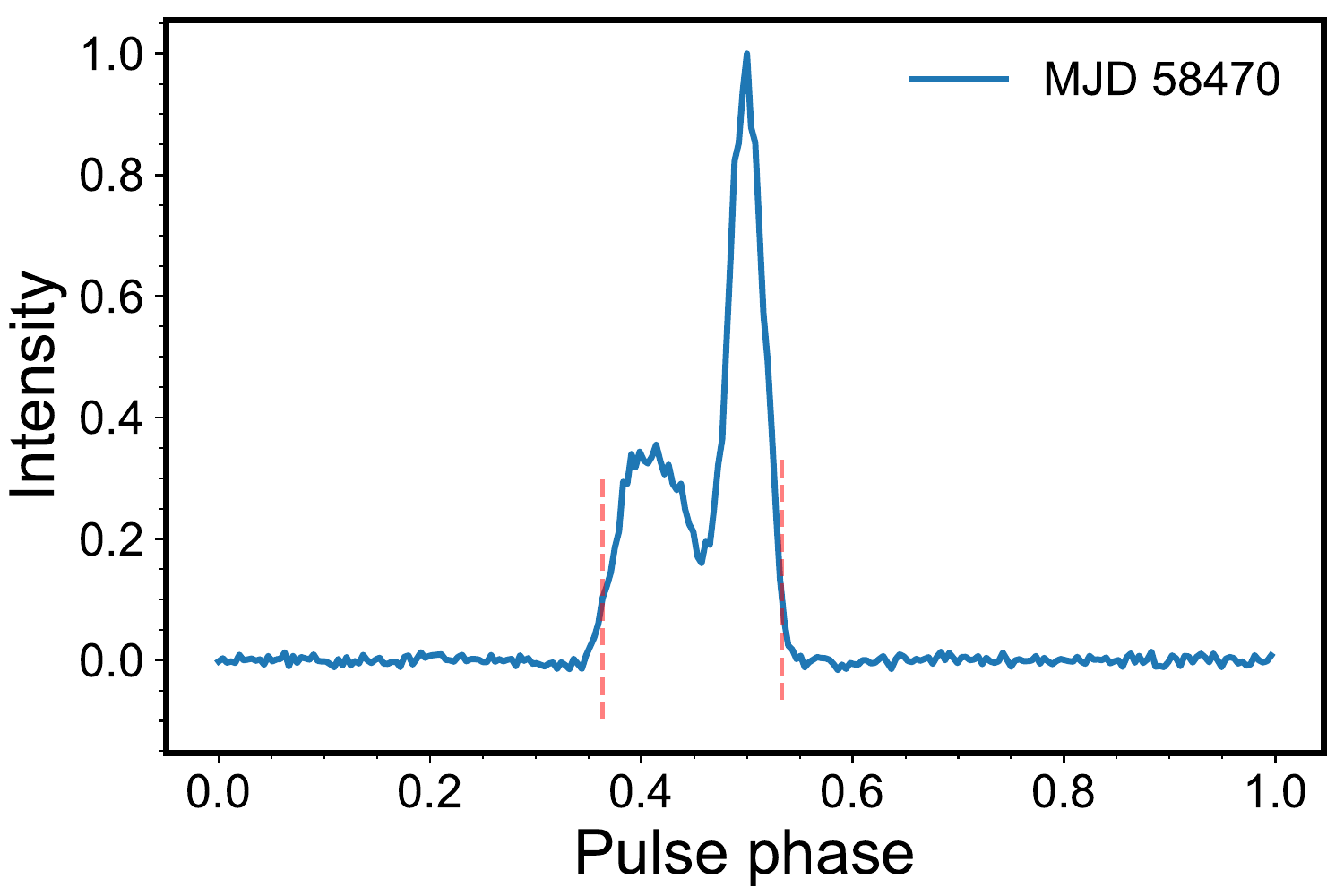}
\label{fig-prof1}}
\subfigure[]{\includegraphics[width=0.3\textwidth]{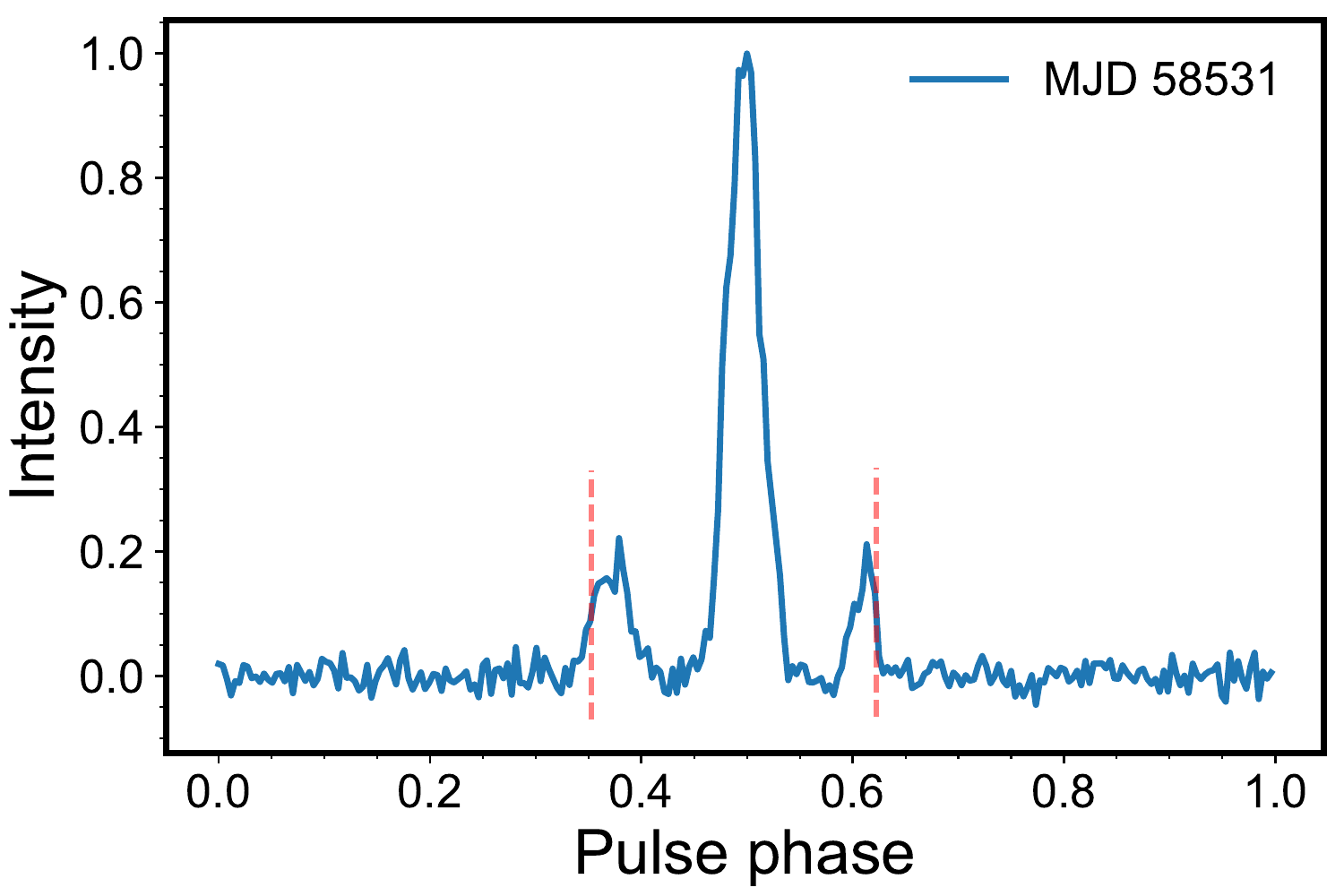}
\label{fig-prof2}}
\subfigure[]{\includegraphics[width=0.3\textwidth]{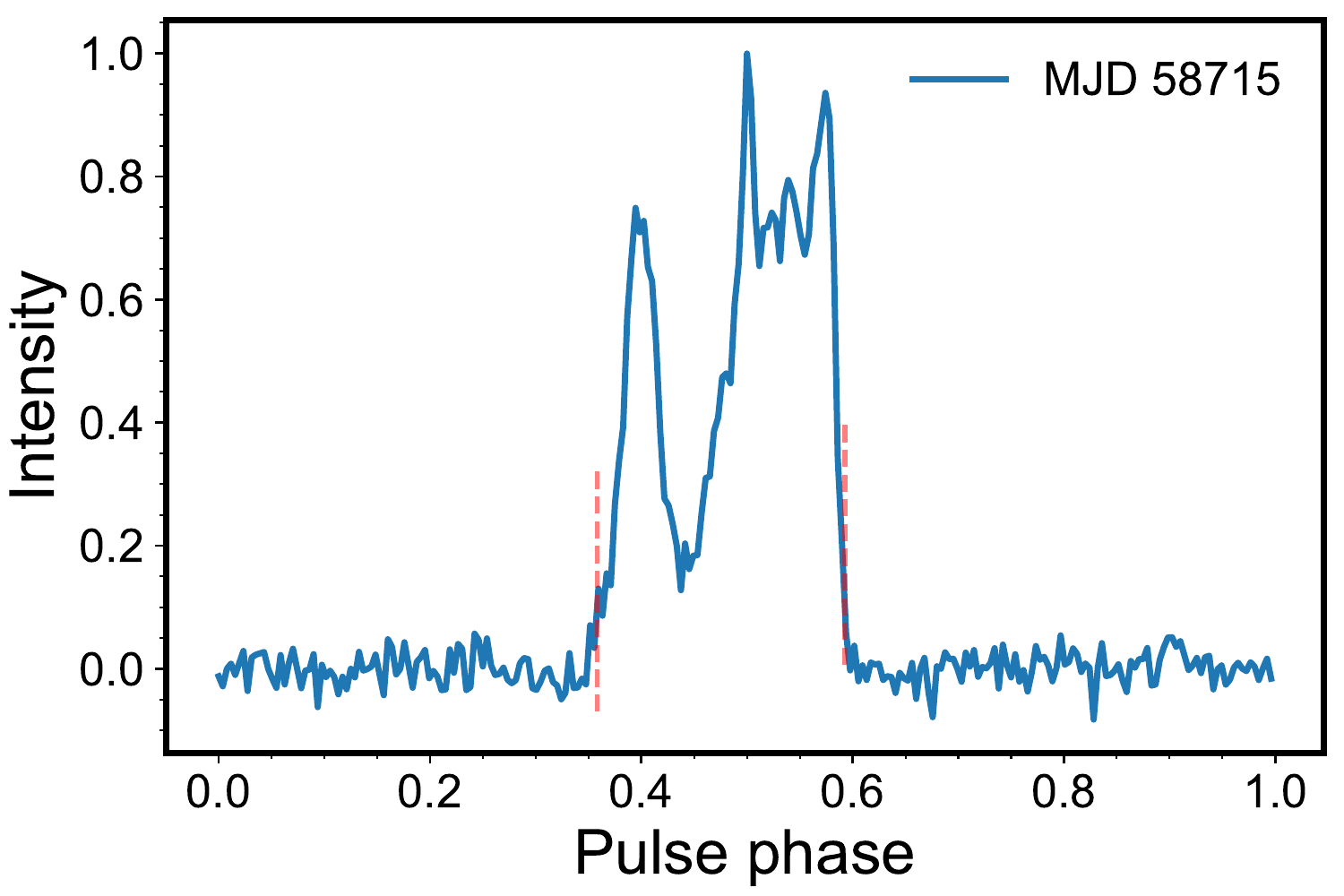}
\label{fig-prof3}}
\subfigure[]{\includegraphics[width=0.3\textwidth]{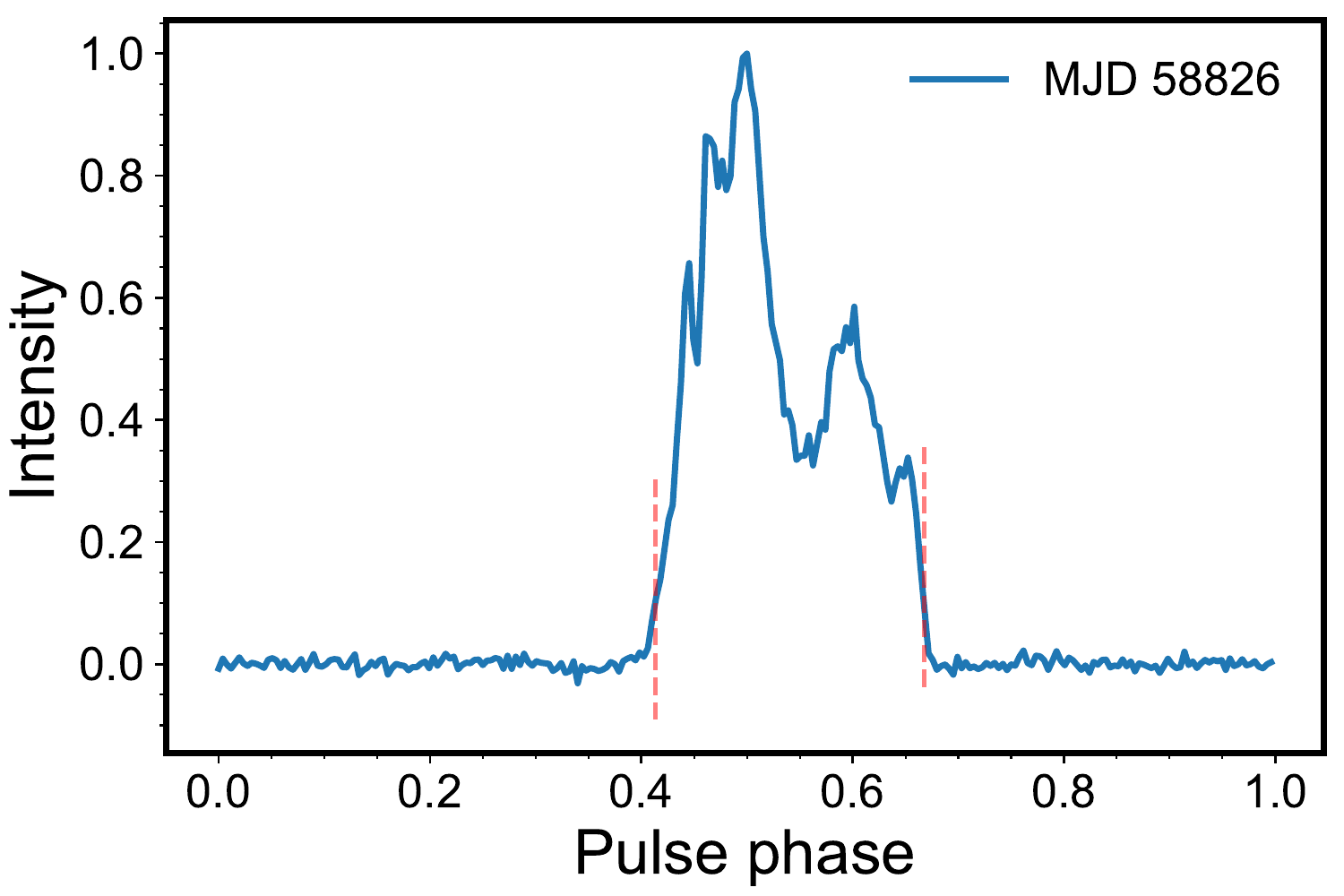}
\label{fig-prof4}}
\subfigure[]{\includegraphics[width=0.3\textwidth]{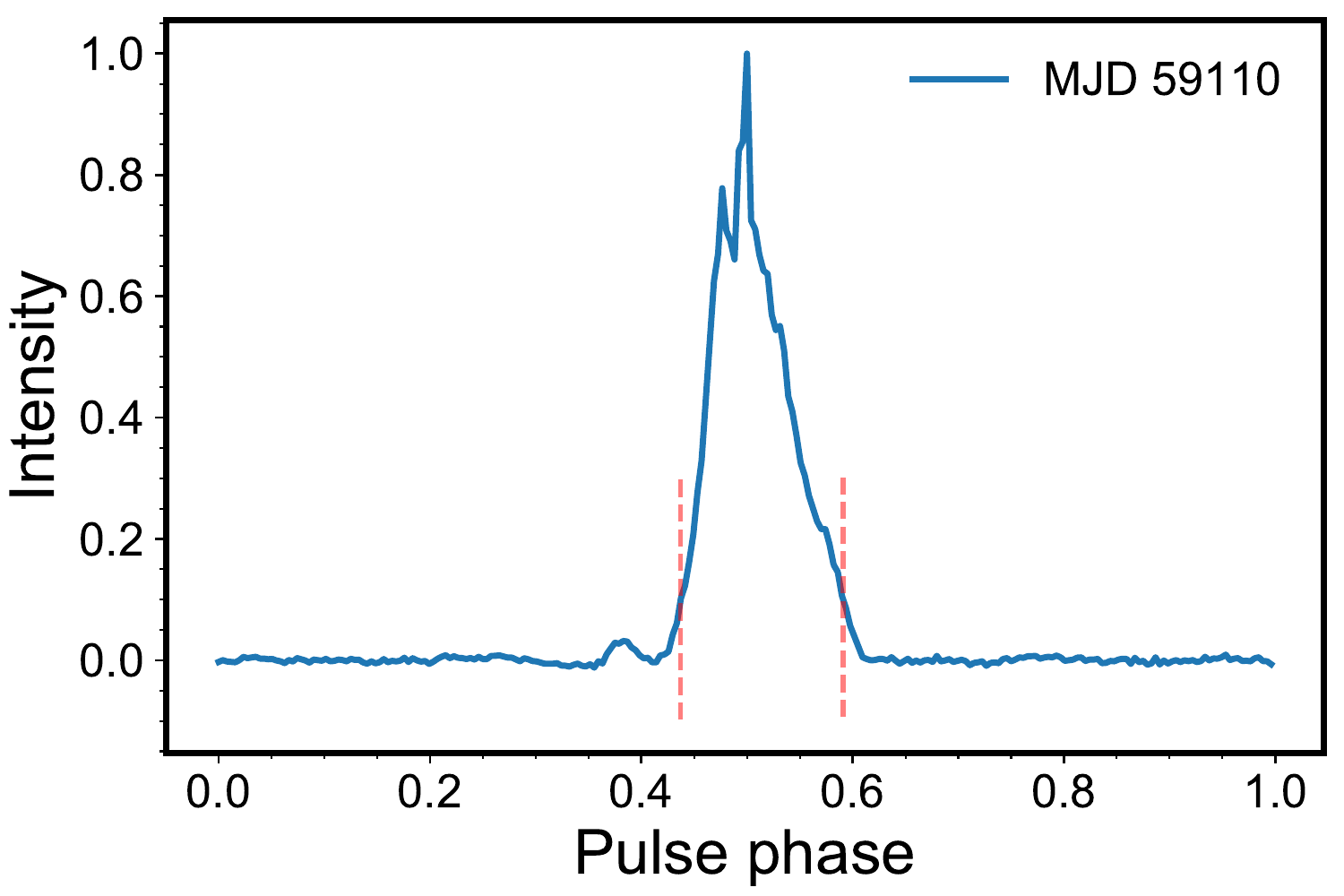}
\label{fig-prof5}}
\subfigure[]{\includegraphics[width=0.3\textwidth]{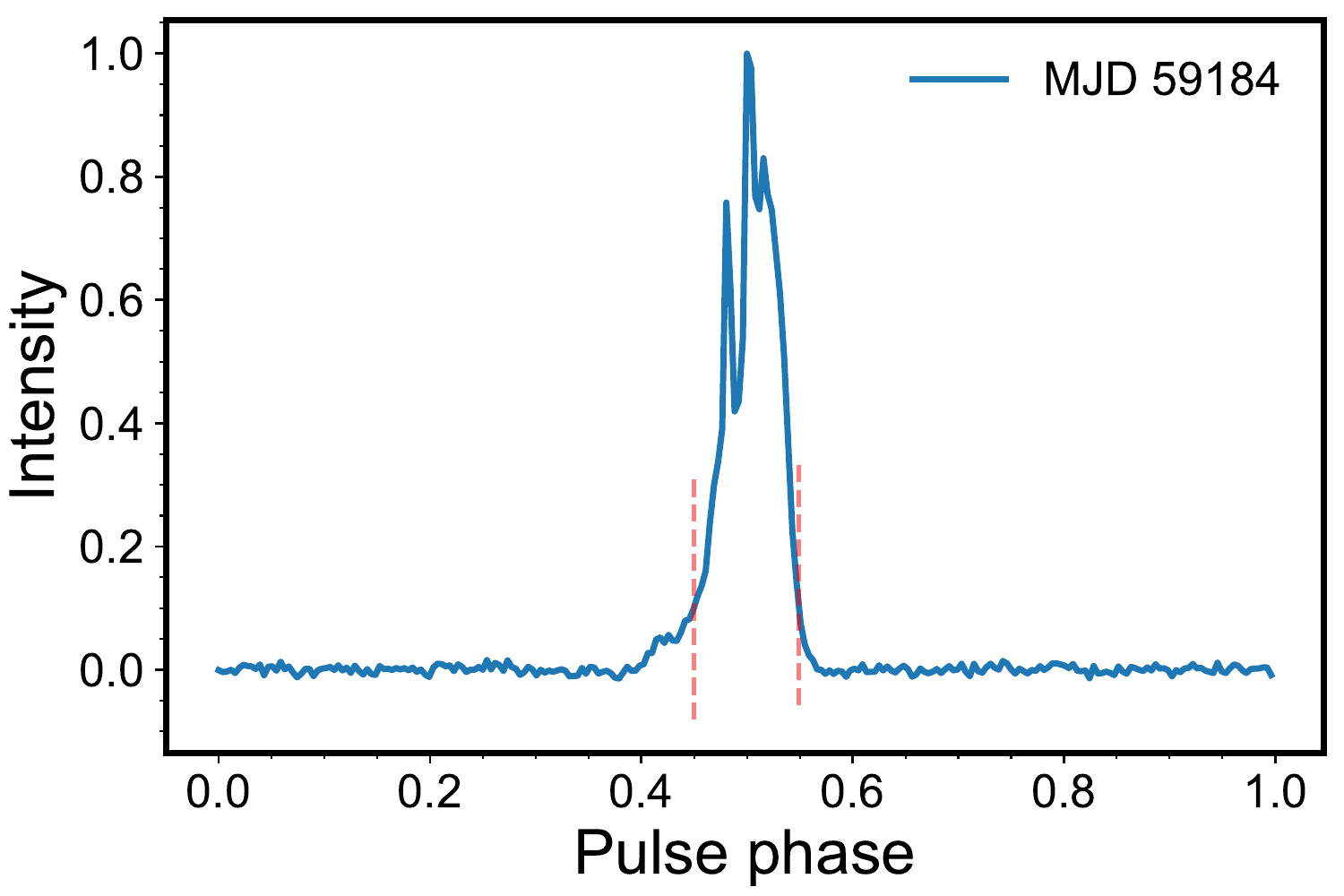}
\label{fig-prof6}}
\subfigure[]{\includegraphics[width=0.65\textwidth]{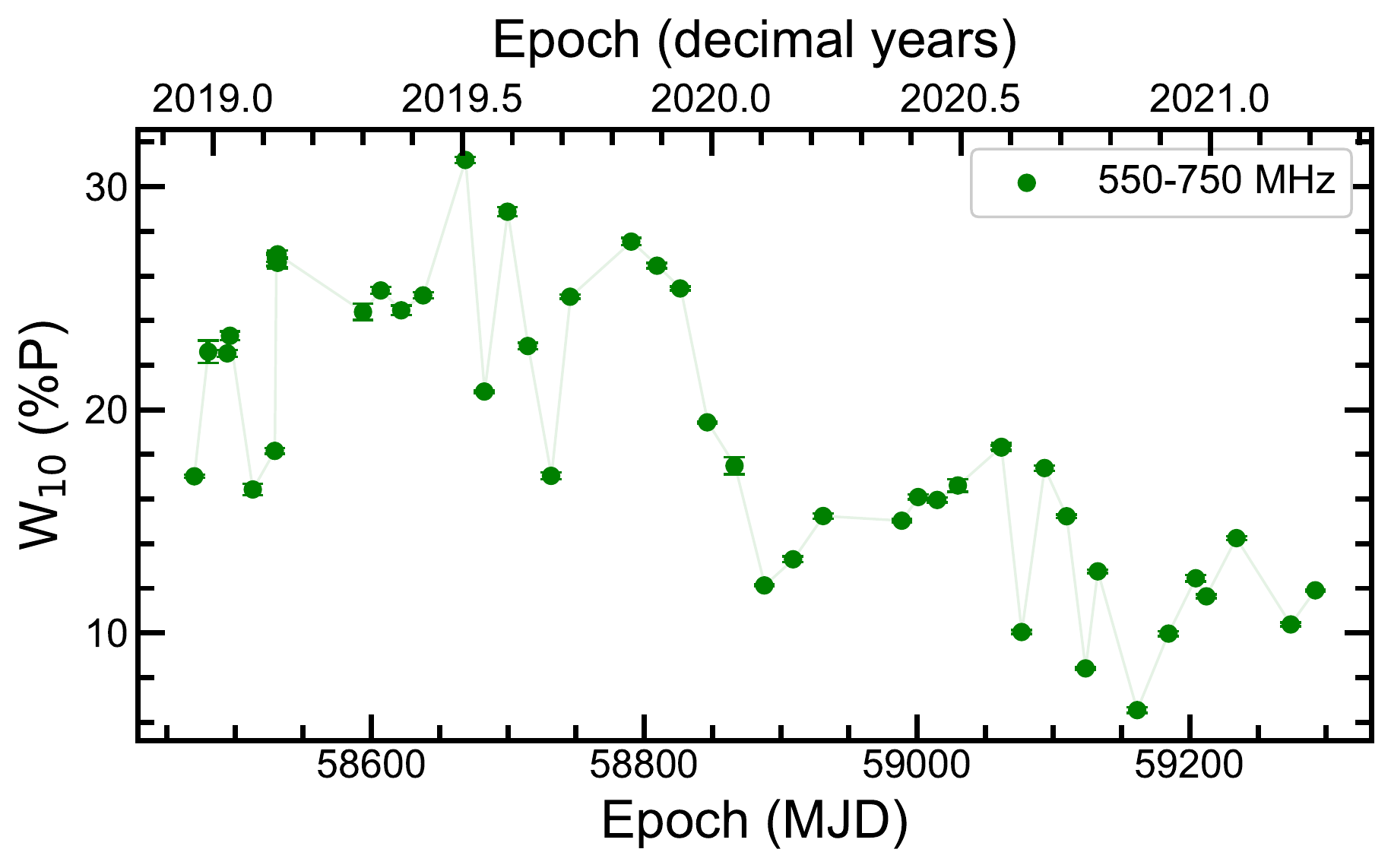}
\label{fig-wd}}
\caption{The panels (a) to (f) show 550$-$750\,MHz (\bandiv) average profiles at six different epochs with their peak intensities normalized to 1. The dashed, red-colored lines in these panels mark the 10\% crossings in the average profiles at the leading as well as the trailing edges, which are used to estimate \wten. The lower panel (g) displays the \wten measured as a percentage of the pulse period for the 550$-$750\,MHz average profiles, as a function of epoch.}
\label{fig-width}
\end{figure*}
\subsection{Spectral index}
The large fractional bandwidths in \bandiii and \bandiv
enable us to estimate the in-band spectral indices. At a given frequency
$\nu$, assuming the flux density, $S_{\nu} \propto \nu^{\alpha}$, we measure
the spectral index $\alpha$ by fitting a straight line in the
$\log S_{\nu} - \log \nu$ plane. The temporal evolution of the in-band
spectral indices measured using \bandiv observations is shown in
Figure~\ref{fig-alpha} (light green points). A general decrease in the
spectral index with time is apparent. Due to poor detection significance
at several epochs and contamination by RFI or baseline variations at several
other epochs, the in-band spectral indices could not be measured reliably
using the band-3 data.
\par
As mentioned in Section~\ref{sec-obs}, a few of our observations were
simultaneous in multiple bands, and several others were
near-simultaneous either in bands 3 and 4 or in bands 4 and 5. The
near-simultaneous observations in two bands were typically separated
by nearly 2\,hours. We can use these simultaneous as well as the
near-simultaneous observations to estimate the spectral indices over
wider frequency ranges, assuming the timescale of the intrinsic flux density
variations of the magnetar is much longer than 2\,hours. The spectral indices
measured using bands 3 and 4, and 4 and 5, covering frequency ranges of
300$-$750\,MHz and 550$-$1450\,MHz, are shown in Figure~\ref{fig-alpha} by red and
blue colored points, respectively. As with the in-band spectral indices
obtained from band$-$4, these inter-band spectral indices also exhibit
a general decreasing trend with time.

\subsection{Profile width}
The average radio profile of \magn changes rapidly with varying number of
components and total shape changes from one epoch to the other.
An example
of this behavior can be seen from the average profiles obtained from
our \bandiv observations at six different epochs that are shown in
Figure~\ref{fig-prof1} to \ref{fig-prof6} \citep[also see Figure 1 in][]{Caleb21}.
To obtain a reasonable estimate of the pulse-width that can be related
to the underlying emission beam, we estimate the positions on the leading
as well as the trailing edge where the intensity crosses the 10\% of the
observed maximum in the profile. In Figure~\ref{fig-prof1} to \ref{fig-prof6},
these crossings are shown using vertical dashed lines in red color.
Using these, we estimate the average profile width at the 10\% level, \wten.
The profile width thus estimated for all the \bandiv profiles which exhibited
a peak S/N of more than 25 is shown as a function of epoch in Figure~\ref{fig-wd}.
While there are significant epoch-to-epoch variations in the profile width,
it is apparent that \wten remains about 25\% of the magnetar's spin period
up to MJD~58800 or so, and then shows a gradual decrease as a function of
time.
\par
We note that the above approach might not take into account very faint leading
or trailing components in the profile. Figure~\ref{fig-prof5} represents one
such example where the peak of a faint pre-cursor component reaches only a few
percent of the maximum in the profile. From manual inspection of the average
profiles, we could identify only four other epochs where the average profile
exhibits very faint (even fainter than the example in Figure~\ref{fig-prof5})
pre-cursor or post-cursor components which are not accounted in \wten estimates.
We believe that such occasional unaccounted components do not alter the long-term
behavior of the profile width apparent in Figure~\ref{fig-width}.

\begin{figure*}
\centering
\subfigure{\includegraphics[width=0.345\textwidth]{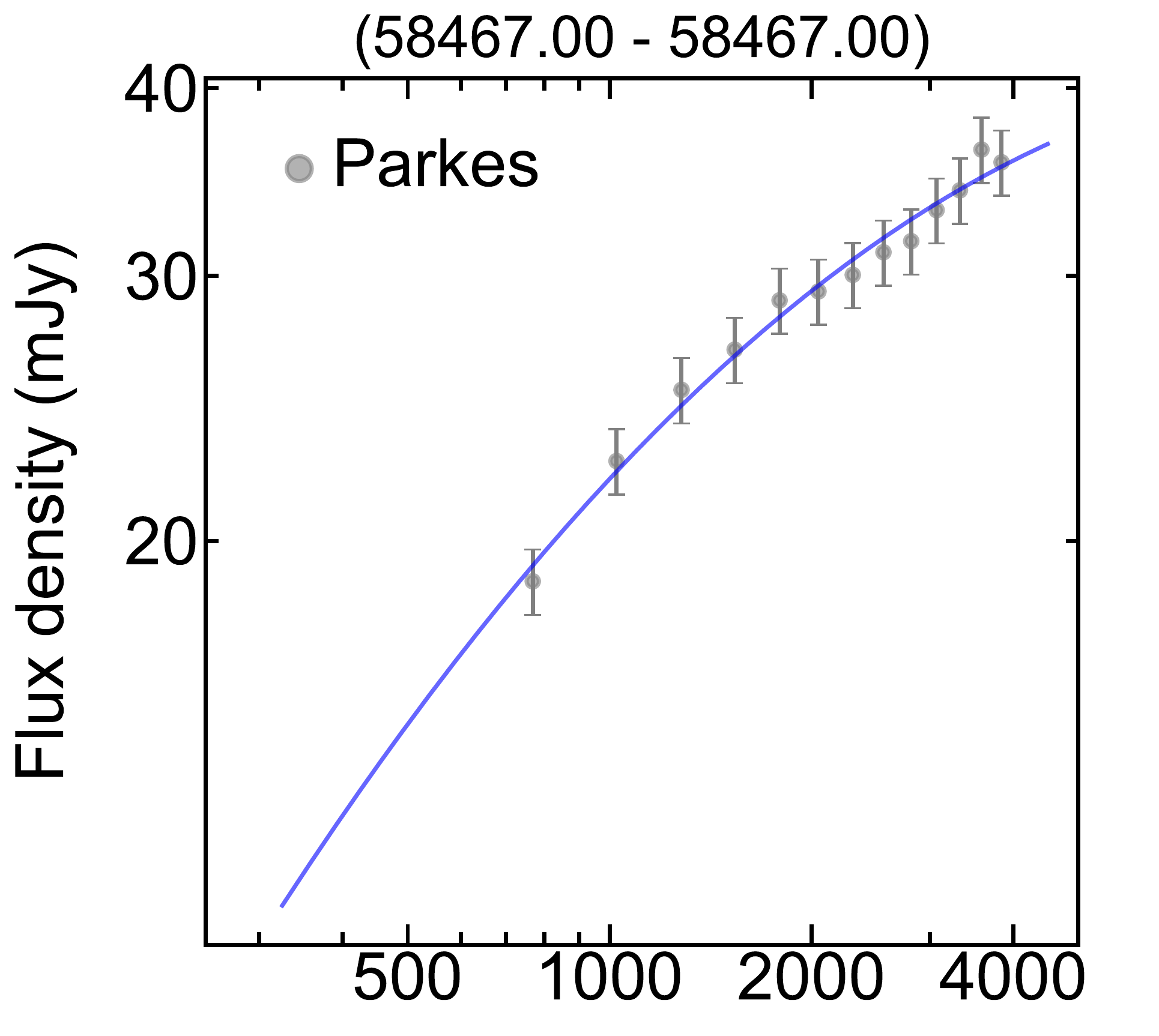}}
\subfigure{\includegraphics[width=0.32\textwidth]{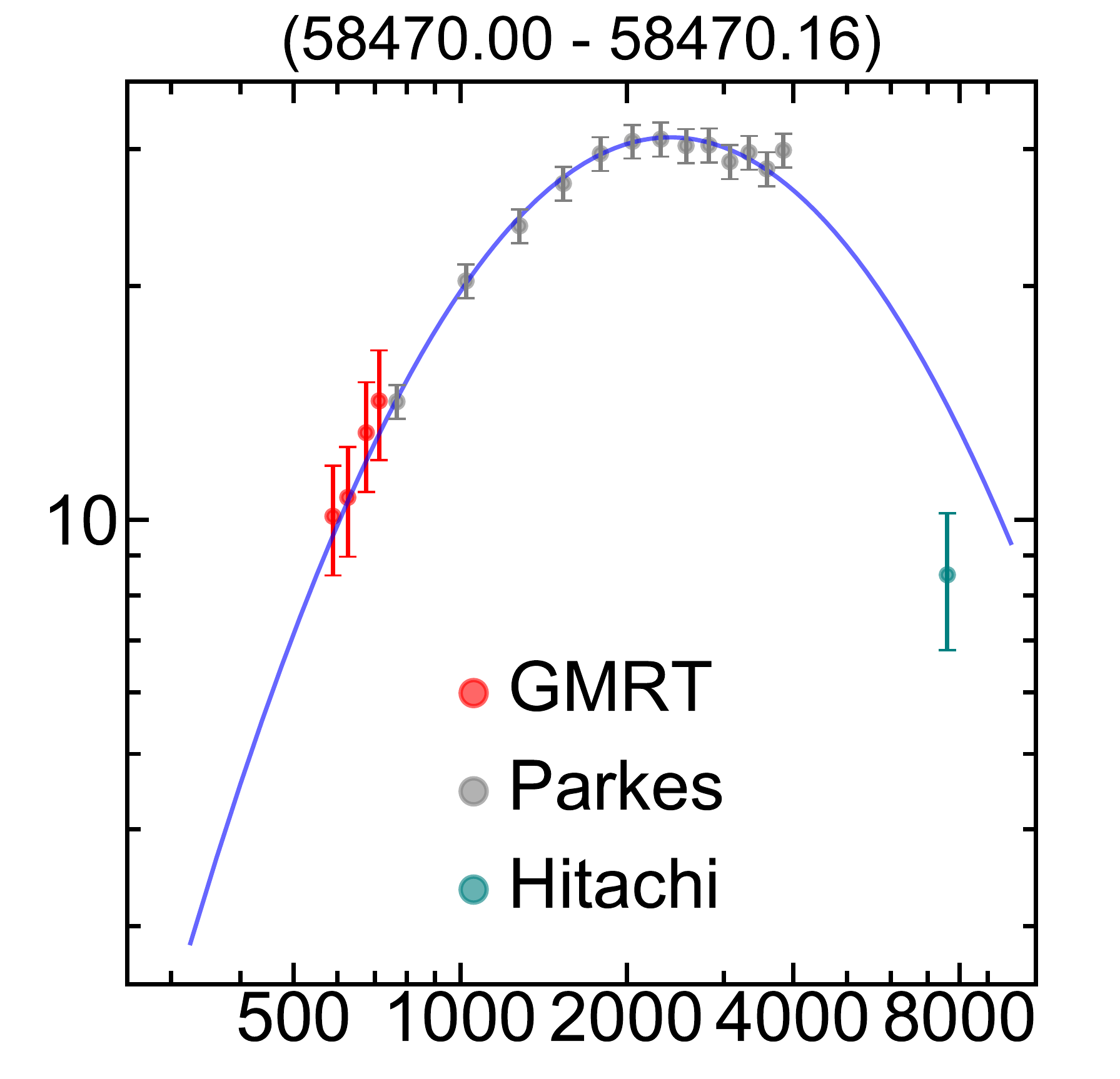}}
\subfigure{\includegraphics[width=0.32\textwidth]{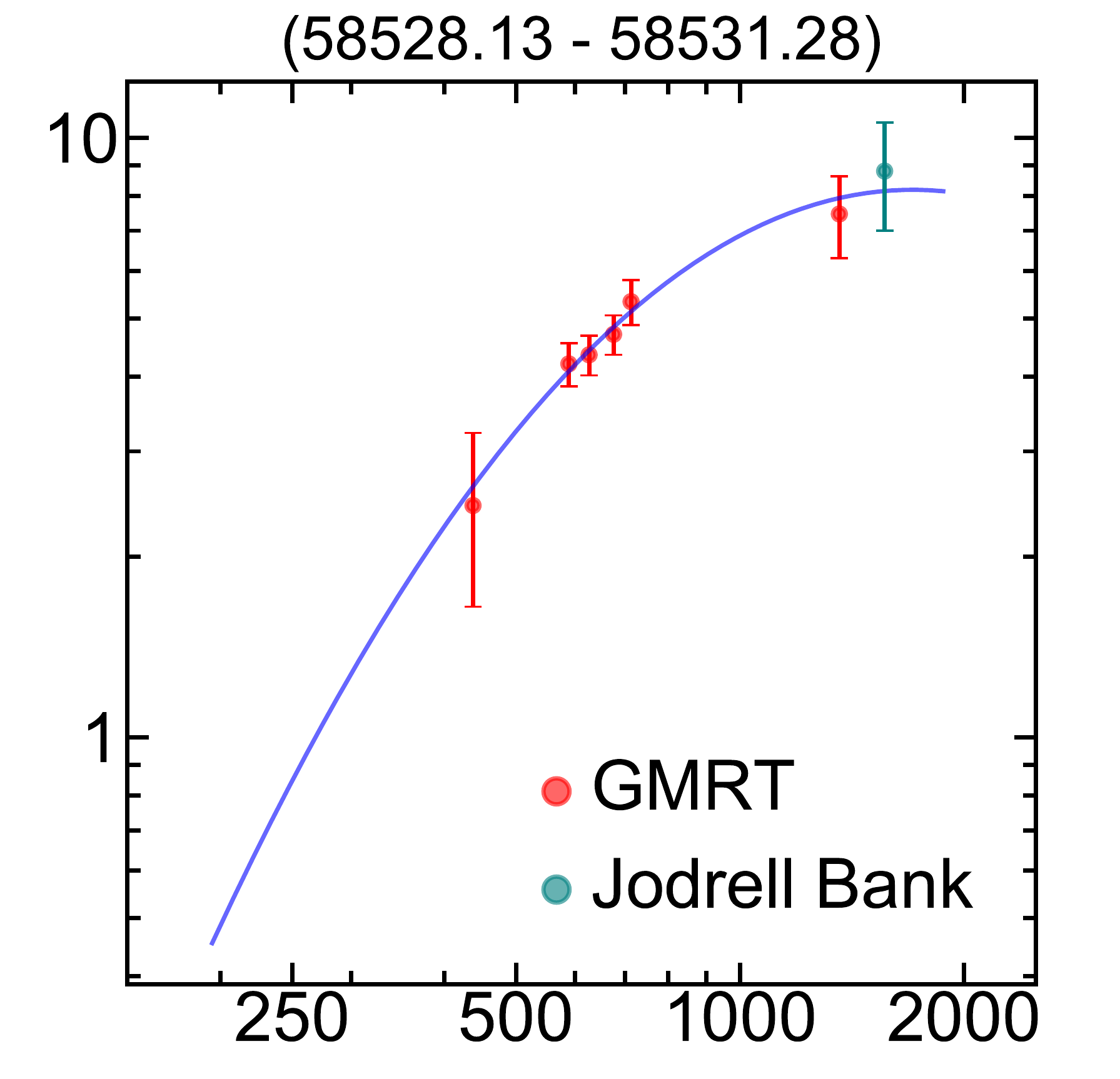}}
\subfigure{\includegraphics[width=0.345\textwidth]{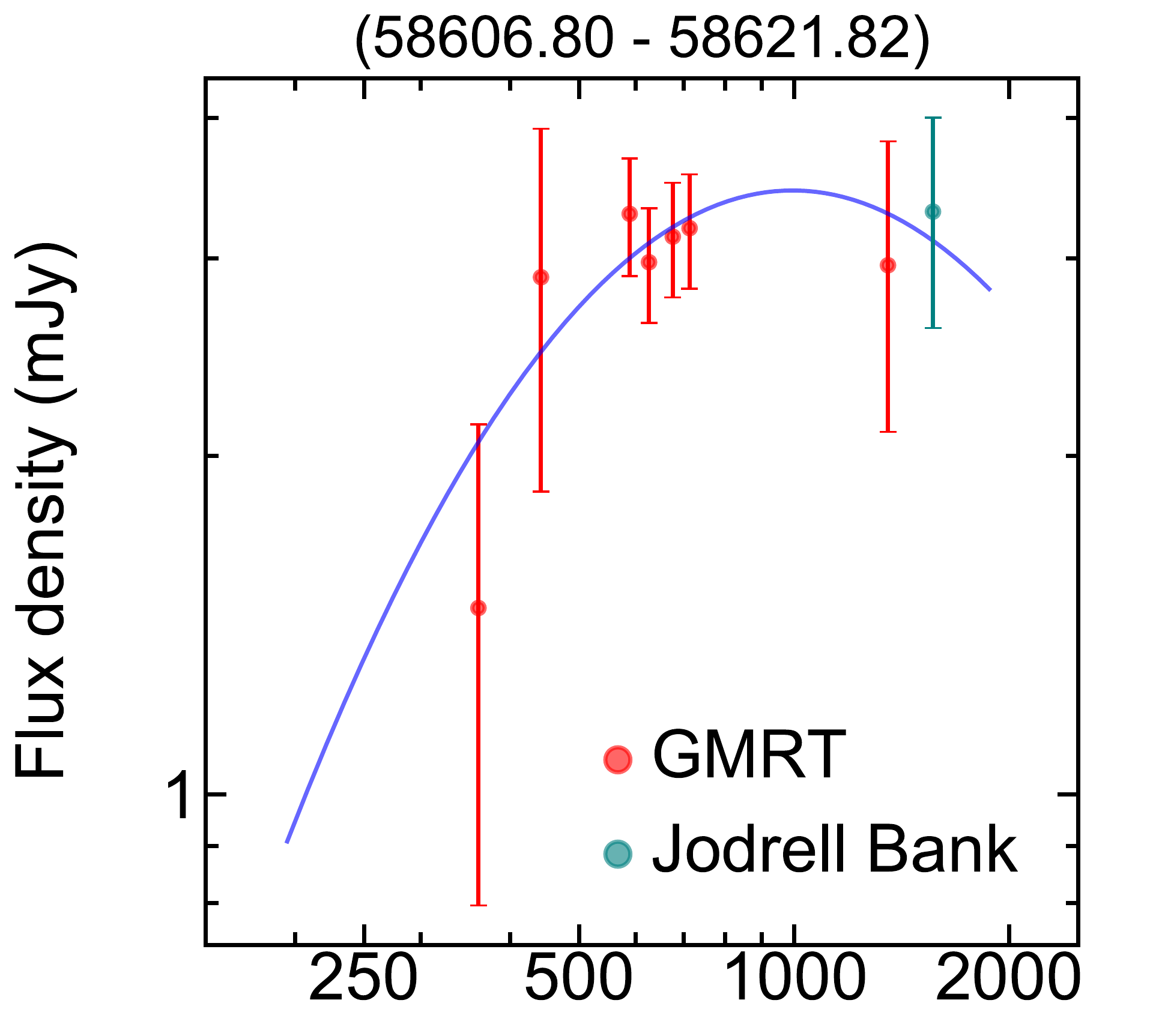}}
\subfigure{\includegraphics[width=0.32\textwidth]{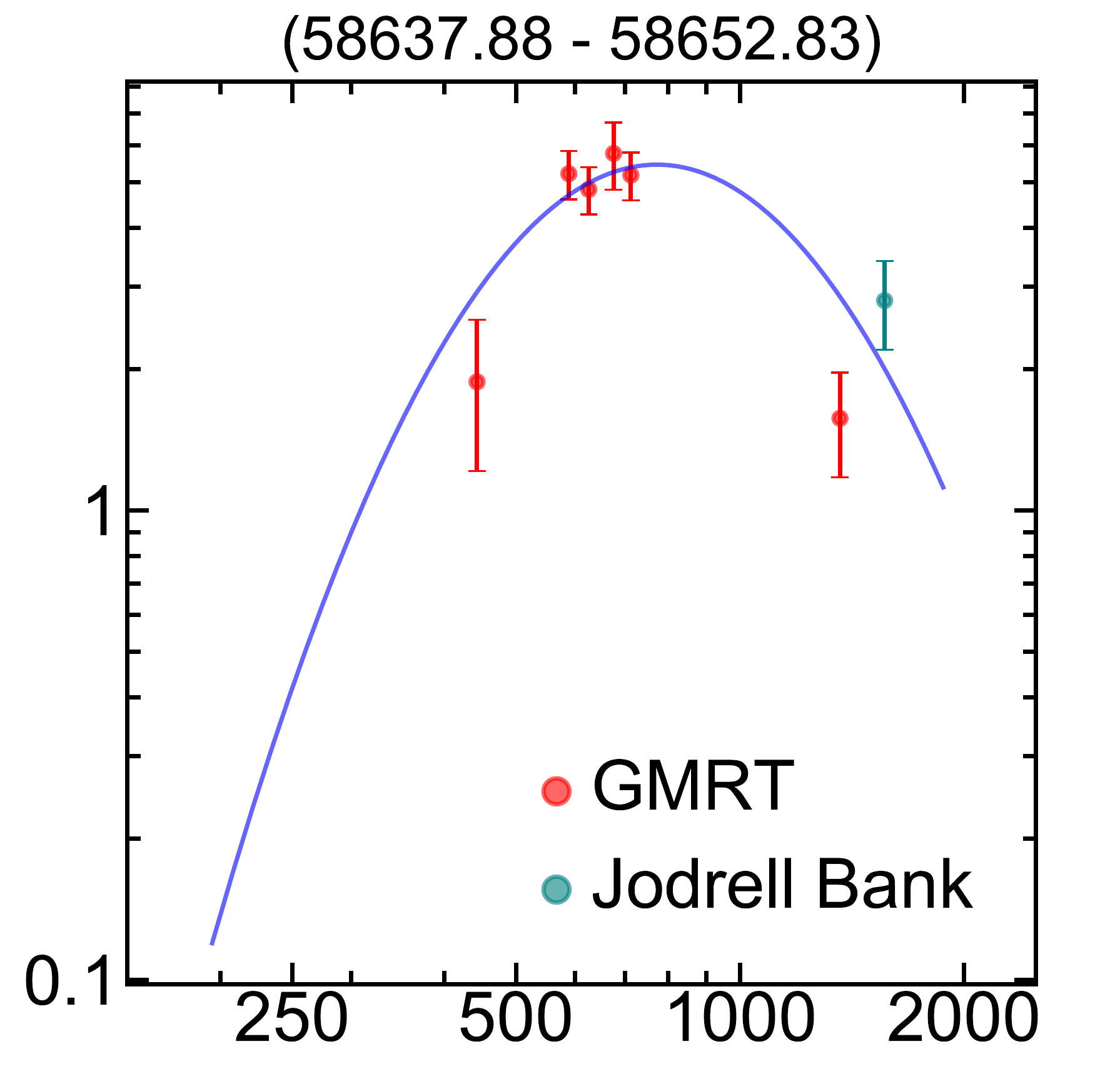}}
\subfigure{\includegraphics[width=0.32\textwidth]{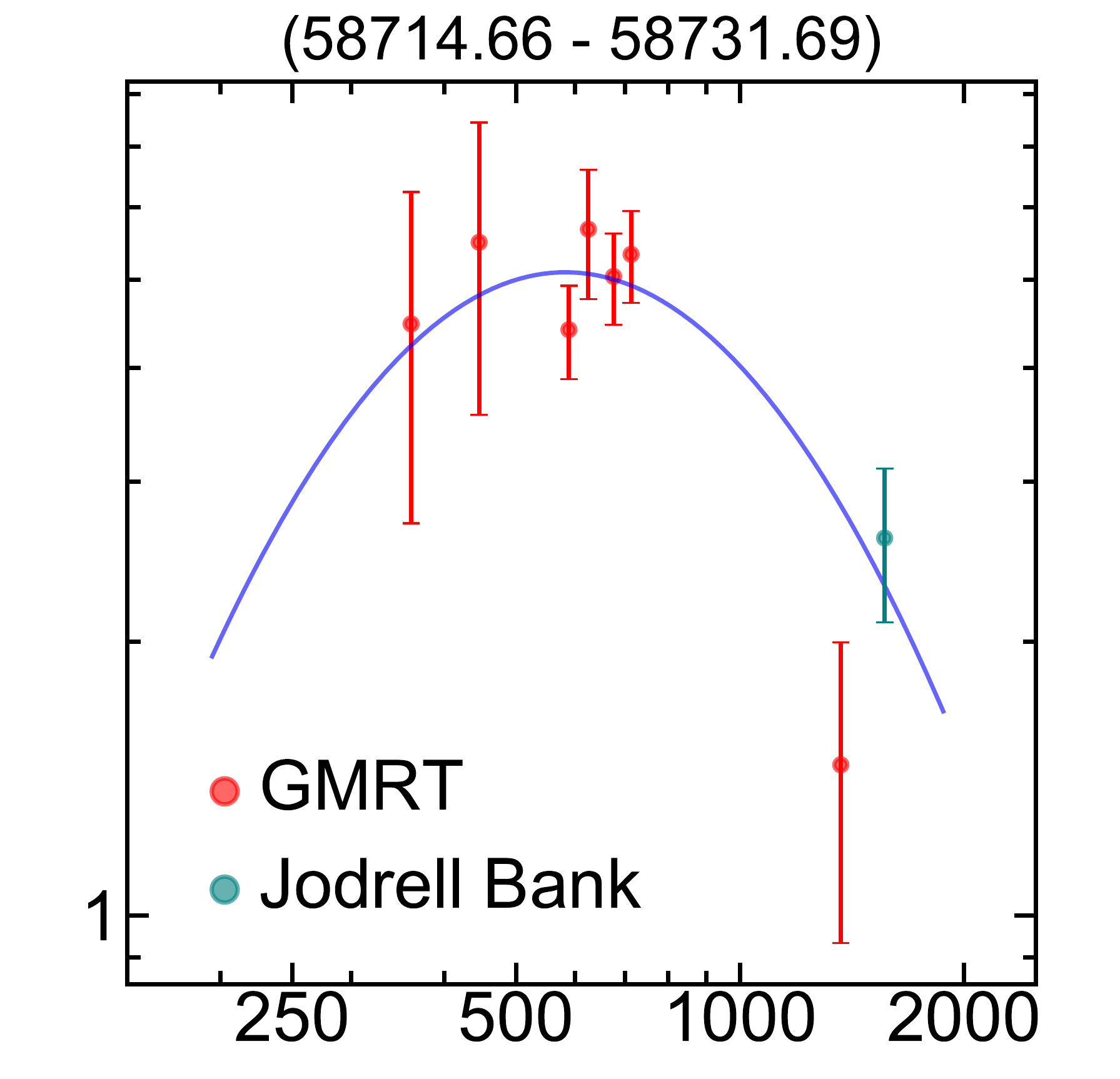}}
\subfigure{\includegraphics[width=0.345\textwidth]{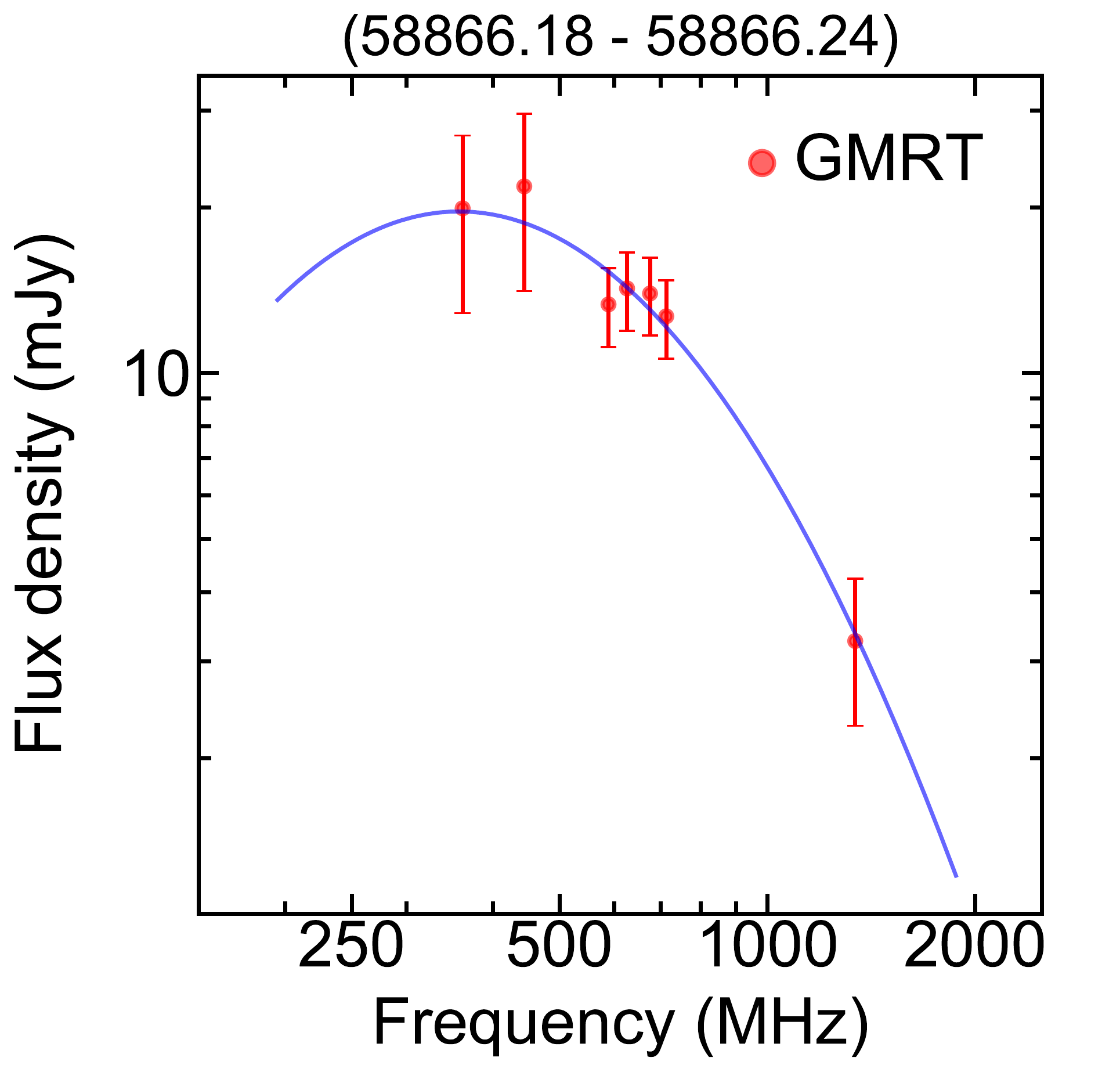}}
\subfigure{\includegraphics[width=0.32\textwidth]{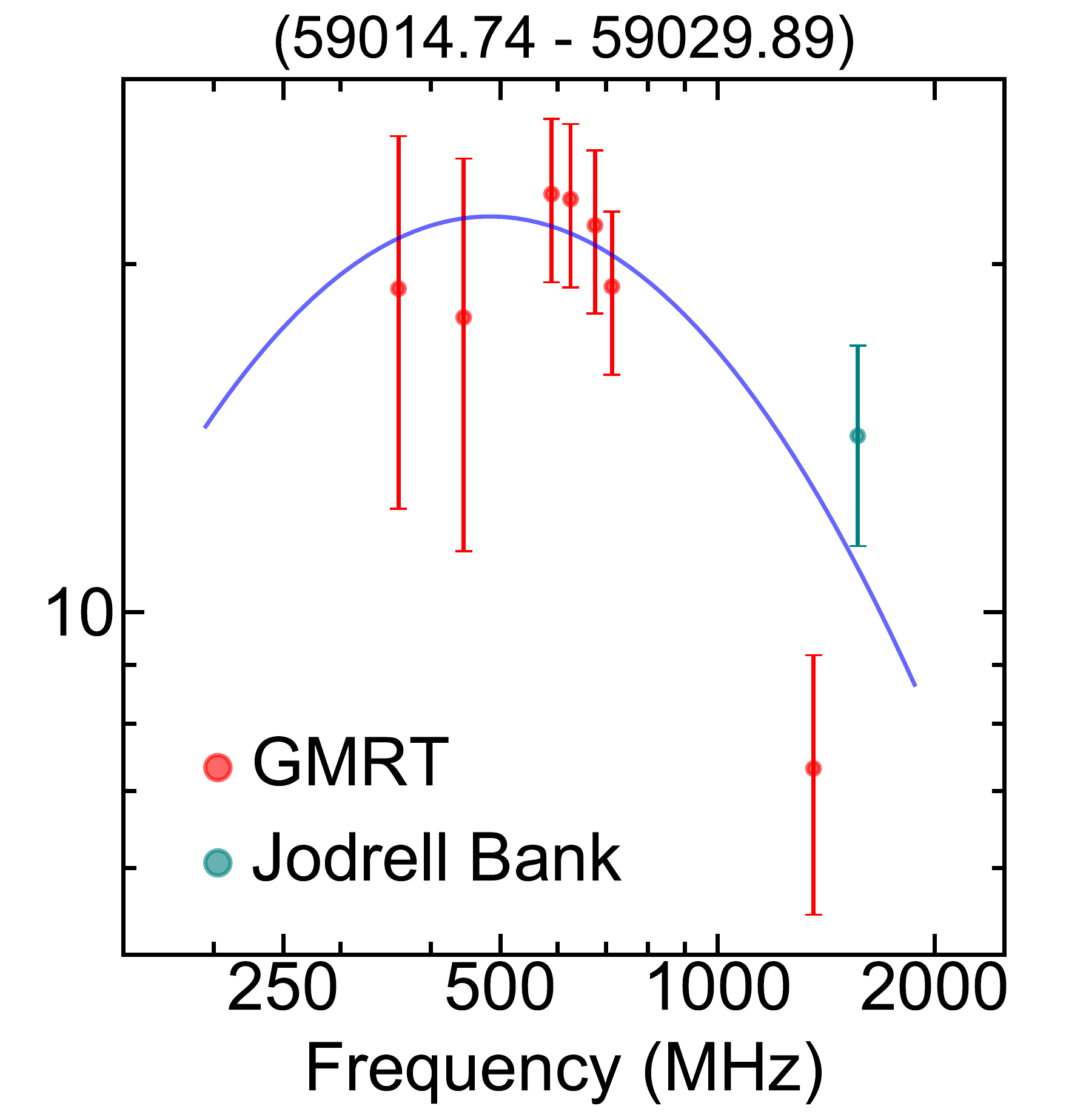}}
\subfigure{\includegraphics[width=0.32\textwidth]{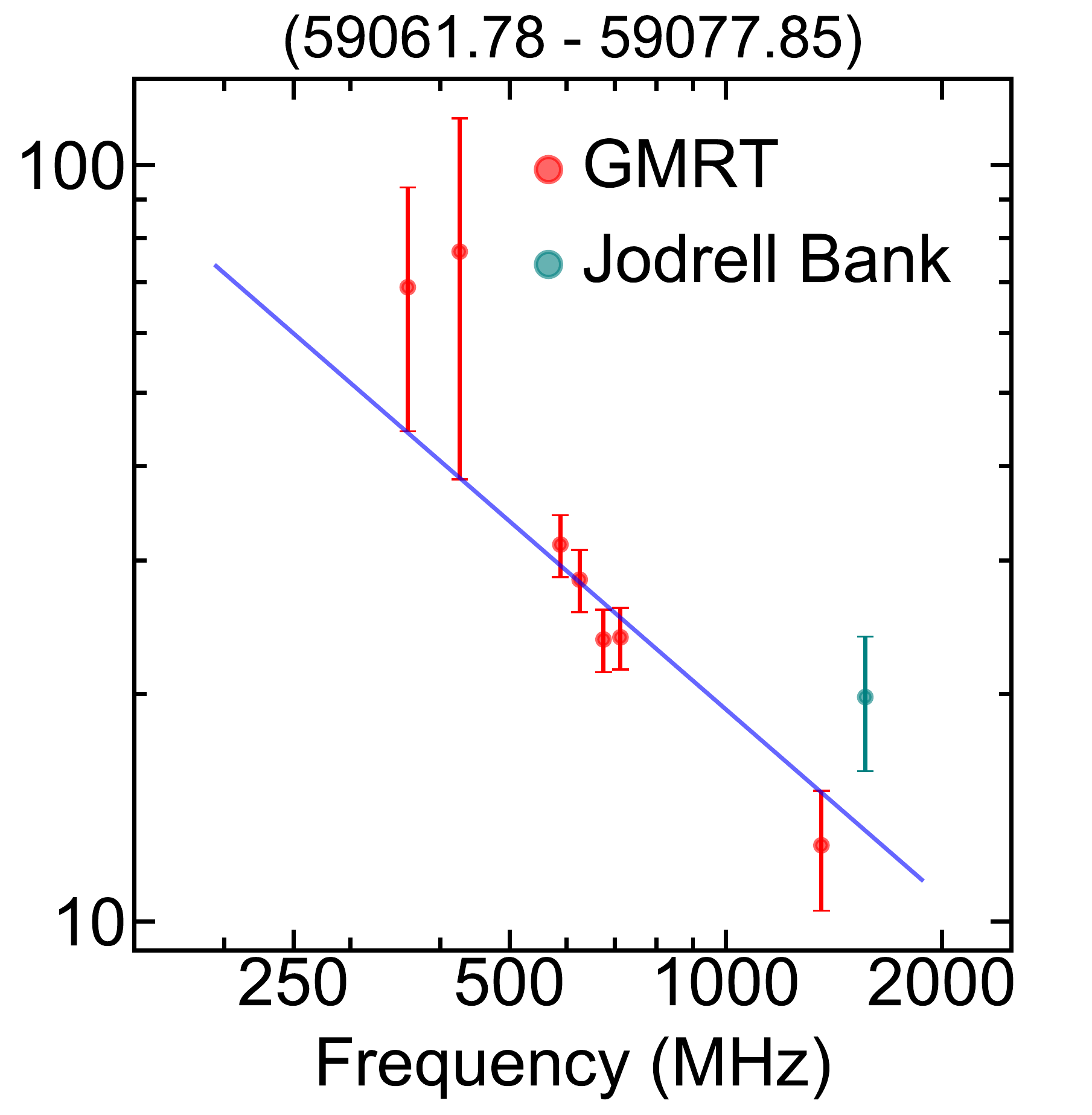}}
\caption{Broadband spectra constructed using multi-band, multi-epoch and
multi-telescope measurement are shown at nine different epochs
(GMRT measurements are from this work, and the Parkes, Hitachi and Jodrell Bank measurements
are from \citet{Dai19}, \citet{Eie21} and \citet{Caleb21}, respectively). Several of
these spectra show clear evidence for a turnover. Also shown are the fits
to these spectra, assuming a log-parabola shape of the spectrum.}
\label{fig-spectra}
\end{figure*}

\section{A temporally evolving, turnover in the spectrum?} \label{sec-turnover}
From the spectral indices shown in Figure~\ref{fig-alpha}, it is evident that
the magnetar's radio spectrum has gradually evolved from inverted spectrum
closer to
the onset of the outburst to as steep as the normal pulsar population at
later epochs. However, Figure~\ref{fig-alpha} also indicates that the
overall spectrum in the lower frequency range (300$-$750~MHz) is flatter
compared to the 550$-$750~MHz range, while that in the higher frequency
range (550$-$1450~MHz) is steeper. This trend of spectral indices measured
from the lower and higher frequency ranges encompassing those measured from
only band~4 continues till about MJD~59000$-$59050, beyond which the spectral
indices measured from all the three frequency ranges seem to follow closely.
\par
The above behavior could potentially be explained by a turnover in the
spectrum with a peak-frequency that evolves downward with time. To test
this hypothesis, we tried to construct broadband spectra at different epochs
using our own observations as well as those available from the literature,
and fit a spectral shape with a turnover. From our own observations, we
consider combining our near-simultaneous observations in bands 3 and 4,
and those in bands 4 and 5, to construct a broadband spectrum covering the
frequency range 300$-$1450~MHz. These two sets of observations were
separated by typically 15~days. As the magnetar exhibits rapid
flux density variations, combining observations from different epochs
might provide an incorrect representation of the overall spectrum.
As \bandiv was common in both the sets of observations, we use the
measured band-4 flux density to gauge if the flux density has evolved or not,
and combine only those sets where the band-4 flux density is relatively
unchanged. This approach resulted in broadband spectra at seven epochs.
\par
From the literature, we looked for the flux density measurements available
during or at nearby epochs for which we could construct the above described
broadband spectra or independent wide-band measurements. We have used published
results from Parkes wide-band
observations at two epochs \citep[December 15 and 18, 2018;][]{Dai19}, 
7.6\,GHz measurement from the Hitachi radio telescope at one epoch \citep{Eie21},
and Jodrell Bank L-band observations at 6 epochs \citep{Caleb21}. Wherever
the uncertainties were less than 5\% or not available at all, we
assumed a uniform 5\% uncertainty on the published flux densities.
The Parkes wide-band measurements resulted in
two additional broadband spectra. The nine broadband spectra that
are constructed this way, including the published measurements, are shown
in Figure~\ref{fig-spectra}.
\begin{figure}
\centering
\includegraphics[width=0.45\textwidth]{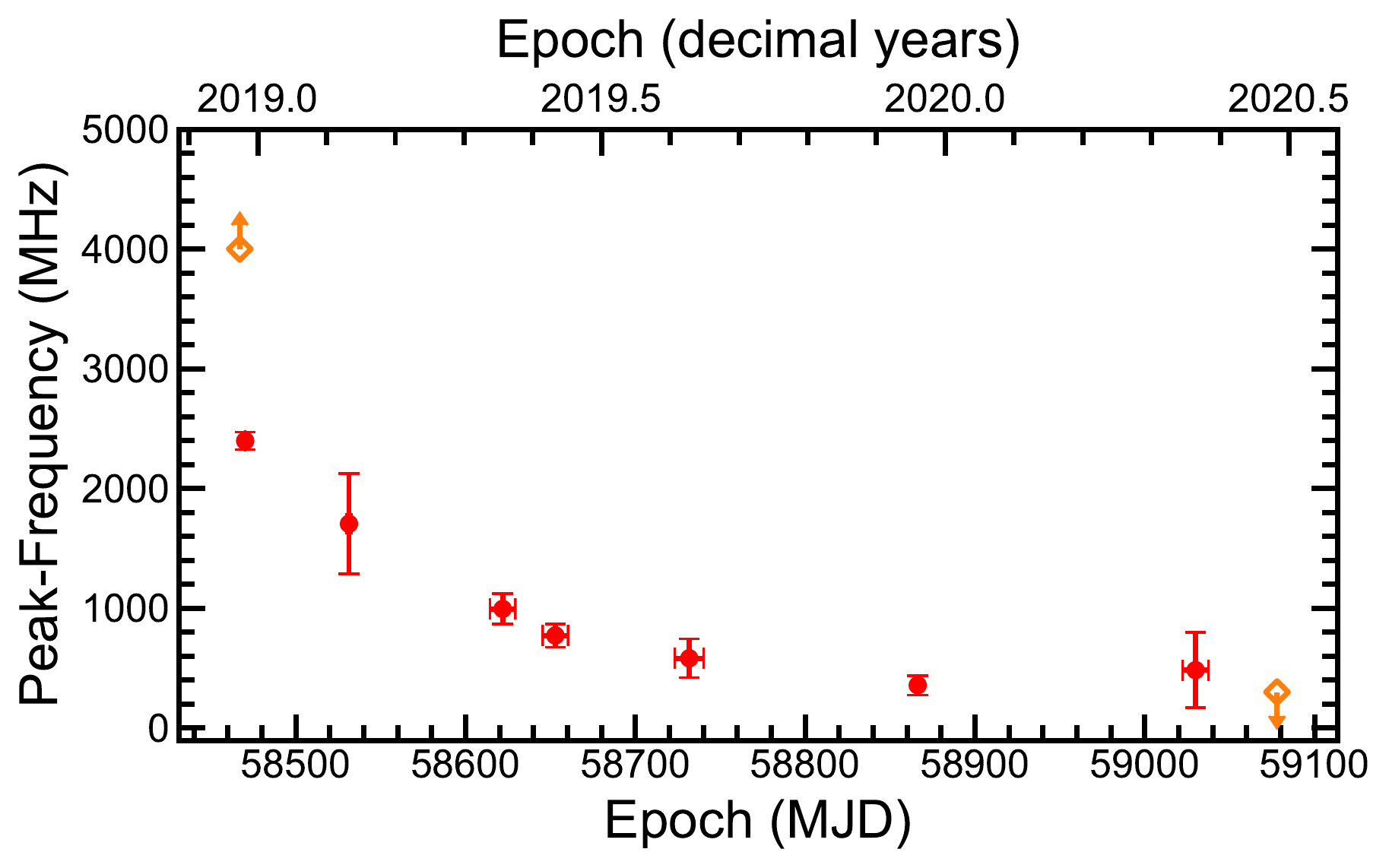}
\caption{The red-colored points show the measured turnover frequency
as a function of epoch. Temporal evolution of the turnover frequency
is clearly evident. The orange colored points indicate rough lower
or upper limits where the turnover frequency could not be constrained.}
\label{fig-tfreq}
\end{figure}
\par
We fit the broadband spectra with a spectral shape of the form of
a log-parabola as following.
\begin{equation}
\logten S_\nu = \beta \times (\logten\nu - \logten\peakf)^2 + \cnot,
\end{equation}
where \peakf is the peak frequency corresponding to the turnover, and
$\beta$ and \cnot are constants. The fitted spectra are also shown
in Figure~\ref{fig-spectra}, overlaid on the data points. The peak
frequencies could not be constrained for the broadband spectrum corresponding
to MJD~58467 (the first subplot in Figure~\ref{fig-spectra}), where it seems
to be inverted in the frequency range that the data points span. Subsequently,
the next four spectra are reasonably well fit by the above log-parabola
model, and the fitted peak-frequency seems to evolve rapidly with time.
The next three spectra are also reasonably well fitted with the above model, 
however, given the spread in our band-3 and band-4 measurements, these
spectra might also be consistent with a flattening at low frequencies rather
than a turnover in the spectrum. In any case, the fitted peak-frequencies
for these three spectra are roughly consistent with each other. For the
last broadband spectrum covering the MJD span 59061.78$-$59077.85 (the last
subplot in Figure~\ref{fig-spectra}), the peak-frequency could not be
constrained and the spectrum appears to be steep throughout the frequency
range covered by the data points.
\par
From the log-parabola fits of the broadband spectra, it is evident that
in about 250\,days following the onset of the outburst, the fitted peak-frequency
evolves quite rapidly. Thereafter, either the peak-frequency appears to settle
down around 500$-$600\,MHz or the spectrum becomes flatter at these low frequencies.
This trend is evident in Figure~\ref{fig-tfreq}. The last broadband spectrum
in Figure~\ref{fig-spectra}, our spectral indices measured using bands 3 and 4
in Figure~\ref{fig-alpha} as well as the flux density measurements shown in
Figure~\ref{fig-flux} suggest that the magnetar exhibits a steep spectrum at
later epochs.
\section{Discussion}
\label{sec-disc}
Following the first high-energy outburst, the magnetar \magn was detected
as a pulsed radio emitter \citep{Camilo06}, and 
subsequently observed successfully at radio frequencies as high as
144\,GHz, infrared as well as X-rays \citep[e.g.,][]{Camilo07c}. Except
for a few observations, the regular monitoring of the magnetar as well as
the measurements of its radio spectrum involved frequencies around and
higher than 1.4\,GHz.
\citet{Camilo07c} monitored the spectrum of the magnetar between
May 2006 and November 2006, at frequencies mostly spanning the range 1.4$-$9\,GHz.
During this time span, the spectrum changed from nearly flat ($\alpha=0$) to
steep ($\alpha=-1$). \citet{Lazaridis08} reported measurements of spectral
indices using simultaneous multi-frequency (1.4$-$14.60\,GHz) observations
between May and July 2006. They showed that overall their measurements were
consistent with those from \citet{Camilo07c}. \citet{Lazaridis08} also
reported spectral index measurements obtained using quasi-simultaneous
multi-frequency (2.64$-$32\,GHz) observations between December 2006 and
July 2007. During this time, they reported the spectral indices which
appear to gradually increase from around $-0.5$ to $+1.0$.
They concluded that overall the spectrum is generally flat,
and there is a slight trend of the spectral index becoming positive with
time. We note that their observations covered different parts of
the above mentioned frequency range at different epochs. \citet{Camilo16}
reported flux densities from their monitoring observations at 1.4\,GHz
and 2\,GHz starting from May 2006 to late 2008 when the radio emission
from the magnetar became undetectable. Although their observations were
not simultaneous, using the 1.4 and 2\,GHz flux density measurements
covering similar time-spans, they asserted that the magnetar's spectrum
was steep between early 2007 and late 2008. Note that \citet{Lazaridis08}
reported the spectrum to become flatter or even inverted till about
mid-2007, however, their measurements covered a different frequency
range.
\par
The above studies provided evidences of significant changes in the
magnetar's radio spectrum during its radio-active phase following the
first outburst. However, due to observations covering different frequency
ranges, and often at different times, it was not possible to study any
underlying gradual evolution of the spectrum with time. Similarly, in
this second, ongoing radio-active phase of this magnetar, \citet{Eie21}
have reported their sparse measurements between December 2018 and
June 2019 at frequencies of 2.3, 6.9, 8.4 and 22\,GHz. Using these
measurements and those available from the literature, they have
indicated that the magnetar's radio spectrum has evolved to become
steep at Gigahertz frequencies.
\par
Following the second outburst,
we have presented our regular monitoring observations
of \magn in the frequency range 300$-$1450\,MHz, along with the
spectral indices measured therefrom. Using these measurements,
it is evident (see Figure~\ref{fig-alpha}) that the magnetar's spectrum
in this frequency range has gradually evolved, from inverted ($\alpha\sim+2$)
closer to the onset of the outburst in December 2018 to steep around
mid-2019, and quite steep ($\alpha\sim-2$) by early 2021. During this time
span, the flux density also shows intriguing trends. After a steep decline
following the outburst, the radio flux density shows at least two episodes
of re-brightening, one around MJD~58800$-$58850 and another around
MJD~59050$-$59150 (Figure~\ref{fig-flux}). There is no hint of enhancement
in the X-ray flux during these radio re-brightening episodes
\citep[see Figure 17 in][]{Caleb21}. Absence of any prominent high-energy
outbursts suggests that although the radio emission appears
following a X-ray outburst, the underlying radio emission processes remain
highly dynamic long after the outburst.
\par
As apparent from Figure~\ref{fig-alpha}, the temporal evolution of spectral
indices measured from different parts of the radio spectrum is slightly
different --- the higher frequency part of the spectrum seems to become
steep earlier than the lower frequency part. A corroborative evidence of
this can also be seen in Figure~\ref{fig-flux} --- peak of a short-lived
enhancement of flux density during roughly MJD~58500$-$58840 shows up
first at \bandiv (550$-$750\,MHz), and then about 20\,days later at \bandiii.
This effect can be explained
if the radio spectrum exhibits a spectral turnover with a peak frequency
that shifts to lower frequencies with time, an inference that was also
discussed by \citet{Eie21}. Using a number of broadband spectra, we have
shown that the magnetar exhibits a fast-evolving turnover in its
radio spectrum. The peak frequency shifted by nearly a factor of 5 (from
$\sim$2.4\,GHz to $\sim$500\,MHz) in less than 250\,days.
\subsection{Is a varying spectral turnover an ubiquitous property of magnetars?}
A number of radio pulsars exhibit a spectral turnover typically around one
or a few GHz \citep{Kijak11a}. Such spectra have
been named as gigahertz-peaked spectra (GPS). Out of the five magnetars that
are known to exhibit radio emission,
the average spectra of
three magnetars, PSR~J1550$-$5418, PSR~J1622$-$4950 and SGR~J1745$-$2900
\citep{Kijak13,Lewandowski15b},
have also shown the characteristic features of GPS. As we have shown,
the magnetar \magn also exhibited a turnover at gigahertz frequencies
close to the onset of its second outburst. Thus, it joins the group
of these GPS pulsars/magnetars. There have been limited broadband radio
monitoring observations of magnetars following their outbursts which
could probe a spectral turnover evolving as systematically as revealed
by our observations. Nevertheless, using archival data at two different
epochs, \citet{Lewandowski15b} have claimed to observe a turnover frequency
that shifts downwards with time
in the spectrum of the Galactic center magnetar SGR~J1745$-$2900. Broadband
radio monitoring of other magnetars following their future outbursts could
reveal if a varying spectral turnover is an ubiquitous property of magnetars.
\subsection{Possible physical reasons for the varying spectral turnover}
The spectral index evolution presented in Figure~\ref{fig-alpha} exhibits
two kinds of variations. First, epoch-to-epoch variations appear to be
stochastic in nature, and might be intrinsic to the emission mechanism.
The other, long-term observed variations in the spectral index
are more gradual. We note that these long-term variations in the spectral
index as well as
peak-frequency are unlikely to be caused by interstellar scintillation.
The diffractive scintillation bandwidth at 650\,MHz is estimated to be
less than or about 1\,kHz, i.e., much smaller than the bandwidths involving
our measurements, in the direction of this source \citep{Maan19b}. The
transition frequency between the strong and weak scattering in the interstellar
medium is estimated to be 50\,GHz, i.e., much higher than the frequencies
involved in this work. We discuss below two physical explanations of the
temporally evolving turnover in the spectrum, and their plausibility.
\subsubsection{Thermal absorption; a potential magnetar wind nebulae?}
The leading explanation for the GPS feature in pulsar/magnetar spectra
is that it originates due to thermal free-free absorption of an otherwise
steep spectral emission by dense, ionized gas regions, either in the
surrounding environment or along the line of sight. The explanation is
motivated by the fact that majority of the GPS sources are located within
ionized environments such as pulsar wind nebulae (PWNe), supernovae remnants (SNRs)
or HII regions that have high electron densities and emission measures.
\par
Magnetars PSR~J1550$-$5418 and PSR~J1622$-$4950 are potentially associated with
SNRs \citep{GG07,Anderson12},
making the thermal absorption as a plausible explanation for their GPS features.
However, for magnetar SGR~J1745$-$2900, there is no known associated SNR
or PWN. For this source, \citet{Lewandowski15b} proposed that the absorption
is caused by the electron material ejected during the outburst, and the ejecta's
expansion with time could explain the observed change in the peak-frequency.
\par
In case of magnetar \magn, we have measured the spectral turnover
at several epochs and the evolution of the turnover frequency with
time is clearly evident. Let us assume a model similar to that proposed
by \citet{Lewandowski15b}, wherein \magn ejected electron material during
its second outburst and the turnover is caused by the thermal absorption
in this material. Assuming local thermal equilibrium conditions, with
quasi-neutral plasma and the electrons that obey the thermal distribution,
the spectral turnover frequency, i.e., the frequency corresponding to an
optical depth of unity, is given by
\begin{equation}
\frac{\peakf}{GHz} = 0.1736\times\frac{T_e}{K}^{-3/4}\sqrt{\frac{EM}{pc\,cm^{-6}} <g_{ff}>},
\end{equation}
where $T_e$ is the electron temperature, $EM$ is the emission measure,
and $g_{ff}$ is a correction factor
\citep[the Gaunt factor, see ][for details]{RW04}. Assuming the correction
factor to be unity, and a uniform density profile, i.e., $EM=N_e^2 \deld$,
where $N_e$ is the electron density and \deld is the thickness of the
absorber, we have
\begin{equation}
\frac{\peakf}{GHz} = 0.1736\times\frac{T_e}{K}^{-3/4}\times\frac{N_e}{cm^{-3}}\sqrt{\frac{\deld}{pc}},
\end{equation}
Note that a decrease in the turnover frequency, as has been observed,
implies either one or a combination of the following: (a) an increase in
the electron temperature of the absorbing medium, (b) a decrease in $N_e$,
(c) a decrease in \deld. It is hard to imagine a way to increase
the absorber's temperature with time, especially as \magn is not known
to be associated with a SNR or PWN. Moreover, the temporal evolution of the
turnover frequency also seem to be linked with that of the outburst. Similar
to the model proposed for J1745$-$2900, we can consider the electron material
ejected during the \magn's outburst and its expansion in a spherical shell to
be the cause of the observed shift in the turnover frequency. We do not try
to constrain the parameters using this model as more observational information
on $T_e$, $N_e$ and \deld is needed. Nevertheless, we note that the observed
turnover frequencies are possible using the parameter values similar to those
considered for J1745$-$2900 by \citet{Lewandowski15b}.
\par
We further note that the absorption considered in the above discussed model
of the ejecta expanding in a spherical shell is thermal absorption by
non-relativistic electrons. However, the electrons in the ejecta from a
magnetar's magnetosphere are expected to be relativistic, which would
significantly decrease the absorption efficiency. Hence, the
relativistic effects need to be incorporated in optical thickness
and absorption efficiency to accurately gauge the likelihood of the
the above model explaining the observed varying turnover in the
magnetar's spectrum.
\subsubsection{Band-limited emission by varying characteristic energy particles?}
The theoretical advances to explain the radio emission characteristics
specifically from magnetars have remained limited. There are models that
propose radio emission from closed field lines  \citep[e.g.,][]{Beloborodov09},
or, much like radio pulsars, from the open field lines \citep[][]{SMG15}.
However, in these models, there are no specific theoretical predictions for
the radio spectrum in general, and an evolving spectral turnover in particular.
\par
There are some key similarities between normal radio pulsars and magnetars,
such as the polarization position angle sweeps that can be modelled by the
rotating vector model \citep{RC69} and highly linearly polarized single
pulses with position angles locally following the mean position angle
traverse \citep[e.g.,][]{Levin12}, which indicate that the underlying
radio emission mechanisms
are similar. The leading radio emission mechanism in pulsars is the
coherent curvature radiation by particle or soliton bunches
\citep[see, e.g.,][]{RS75,MGP00,MGM09}. Following \citet{RS75}, the
characteristic frequency of single-particle curvature radiation is
given by
\begin{equation}
f_c = \frac{3c}{2\pi\rho} \gamma^3,
\end{equation}
where $c$ is the speed of light, $\rho$ is the radius of curvature at
the emission-site and $\gamma$ is the Lorentz factor of the particle.
The exact shape of the \emph{observed} spectrum is decided by several
factors, such as the energy distribution of the particles producing the
observed coherent radiation via bunches, the viewing geometry and
the range of emission heights, among others.
\par
Here we propose a hypothesis that observed radio emission from the magnetar
\magn is caused by underlying particles with a narrow energy distribution,
and the observed peak in the radio spectrum of the magnetar \magn
corresponds to the characteristic frequency $f_c$. In this highly simplified
picture, the observed downward shift in the peak-frequency can be interpreted
as a corresponding decrease in the energy of the particle population that gives
rise to the observed radiation. As $f_c\propto\gamma^3$, the observed shift of
the peak-frequency by a factor of 5$-$6 needs the population energy to decrease
by a factor less than 2.
\subsection{Profile width evolution}
As apparent from Figure~\ref{fig-width}, the magnetar's average profile width
shows interesting evolution. Apart of significant short term variations, a trend
in the long term is visible. After the outburst, the profile width seems to remain
around 20$-$25\% of the magnetar's spin period, albeit with large fluctuations,
during the first 350\,days or so, until around MJD~58820. Afterwards, the profile
width decreases gradually, becoming around 10\% of the period in March 2021.
\par
There are expected profile width variations in some magnetar emission models.
\citet{Beloborodov09} proposed a model wherein, following an outburst, the magnetosphere gradually untwists and gives rise to non-thermal radiations preferentially generated on a bundle of extended closed magnetic field lines near the dipole axis. In this model, the radio luminosity as well as the pulse-width is expected to decrease as the bundle shrinks in such a way that most of the radio emission is absorbed within the magnetospheric plasma.
While the observed decrease in the pulse-width, particularly at later epochs, seems to be consistent with this picture, we do not see a monotonic decrease in the radio luminosity.
\citet{SMG15} suggest the radio emission from magnetars to originate, much like from normal radio pulsars, from the open field line regions, and explain the emission with the partially screened gap model. However, for the radio emission to be visible, they rely on alteration of the open field line region at the time of outburst to widen the radio beam which causes the radio detection of magnetars in their model. At the time of the outburst, the curvature of open field lines changes significantly, resulting in a much larger opening angle of radio emission. While returning to the quiescent state, the radius of curvature increases back to its original value causing a gradual narrowing of the radio beam, and hence, narrowing of the observed profile width and eventual disappearance of the magnetar. As the predicted behavior is similar in both the models, the observed trend in the profile width does not discriminate between these models.
\par
Overall, the intriguing spectro-temporal evolution uncovered by our
low-frequency monitoring
of the magnetar \magn will hopefully motivate more detailed or even new
theoretical modelling of magnetar radio emission. An evolving spectral turnover
might be an ubiquitous property of radio magnetars, and our findings strongly
advocate systematic radio monitoring of magnetars following their outbursts,
preferably using instruments which offer ultra-wide frequency coverages
\citep[e.g.,][]{Maan13,Hobbs20}.

\section{Conclusions}
We have presented results from a multi-frequency monitoring campaign of the
magnetar \magn (\mpsr) with the GMRT covering a frequency range of
300$-$1450 MHz. Based on the flux density measurements at multiple frequencies,
we see that the flux density of the magnetar has significantly varied over
time at all the frequencies, with multiple episodes of enhanced radio activity.
The width of the 550$-$750\,MHz average profile shows curious behavior: it remains
roughly same for the first 350\,days or so, and gradually decreases afterwards.
A simple power-law modeling suggests that the magnetar's radio spectrum has
evolved from a flatter or even inverted (magnetar-like) to a steeper (pulsar-like)
spectrum with time. A more detailed analysis using broadband spectra reveals
an evolving turnover in the spectrum, with the turnover frequency decreasing
as a function of time. We propose that the thermal absorption by a piece of the
intervening medium, such as the expanding ejecta from the outburst, or
a change in the intrinsic emission, such as a hypothetically decreasing
energy of the leptons generating the curvature radiation in the magnetosphere,
remain plausible physical explanations for the observed spectral behavior.
\acknowledgments
YM thanks Dipanjan Mitra for scientific discussions related to the
work presented in this paper.
We thank the staff of the GMRT who have made these observations possible. The
GMRT is run by the National Centre for Radio Astrophysics of the Tata
Institute of Fundamental Research.
\software{RFIClean \citep{MvLV21}, PRESTO \citep{RansomThesis}, SIGPROC, DSPSR \citep{vSB11}}
\facility{GMRT(GWB)}


\bibliography{refs}{}

\begin{thebibliography}{}
\expandafter\ifx\csname natexlab\endcsname\relax\def\natexlab#1{#1}\fi
\providecommand{\url}[1]{\href{#1}{#1}}
\providecommand{\dodoi}[1]{doi:~\href{http://doi.org/#1}{\nolinkurl{#1}}}
\providecommand{\doeprint}[1]{\href{http://ascl.net/#1}{\nolinkurl{http://ascl.net/#1}}}
\providecommand{\doarXiv}[1]{\href{https://arxiv.org/abs/#1}{\nolinkurl{https://arxiv.org/abs/#1}}}

\bibitem[{{Anderson} {et~al.}(2012){Anderson}, {Gaensler}, {Slane}, {Rea},
  {Kaplan}, {Posselt}, {Levin}, {Johnston}, {Murray}, {Brogan}, {Bailes},
  {Bates}, {Benjamin}, {Bhat}, {Burgay}, {Burke-Spolaor}, {Chakrabarty},
  {D'Amico}, {Drake}, {Esposito}, {Grindlay}, {Hong}, {Israel}, {Keith},
  {Kramer}, {Lazio}, {Lee}, {Mauerhan}, {Milia}, {Possenti}, {Stappers}, \&
  {Steeghs}}]{Anderson12}
{Anderson}, G.~E., {Gaensler}, B.~M., {Slane}, P.~O., {et~al.} 2012, \apj, 751,
  53, \dodoi{10.1088/0004-637X/751/1/53}

\bibitem[{{Beloborodov}(2009)}]{Beloborodov09}
{Beloborodov}, A.~M. 2009, ApJ, 703, 1044, \dodoi{10.1088/0004-637X/703/1/1044}

\bibitem[{{Caleb} {et~al.}(2021){Caleb}, {Rajwade}, {Desvignes}, {Stappers},
  {Lyne}, {Weltevrede}, {Kramer}, {Levin}, \& {Surnis}}]{Caleb21}
{Caleb}, M., {Rajwade}, K., {Desvignes}, G., {et~al.} 2021, \mnras,
  \dodoi{10.1093/mnras/stab3223}

\bibitem[{{Camilo} {et~al.}(2006){Camilo}, {Ransom}, {Halpern}, {Reynolds},
  {Helfand}, {Zimmerman}, \& {Sarkissian}}]{Camilo06}
{Camilo}, F., {Ransom}, S.~M., {Halpern}, J.~P., {et~al.} 2006, Nature, 442,
  892, \dodoi{10.1038/nature04986}

\bibitem[{{Camilo} {et~al.}(2007{\natexlab{a}}){Camilo}, {Reynolds},
  {Johnston}, {Halpern}, {Ransom}, \& {van Straten}}]{Camilo07a}
{Camilo}, F., {Reynolds}, J., {Johnston}, S., {et~al.} 2007{\natexlab{a}},
  ApJL, 659, L37, \dodoi{10.1086/516630}

\bibitem[{{Camilo} {et~al.}(2007{\natexlab{b}}){Camilo}, {Cognard}, {Ransom},
  {Halpern}, {Reynolds}, {Zimmerman}, {Gotthelf}, {Helfand}, {Demorest},
  {Theureau}, \& {Backer}}]{Camilo07b}
{Camilo}, F., {Cognard}, I., {Ransom}, S.~M., {et~al.} 2007{\natexlab{b}}, ApJ,
  663, 497, \dodoi{10.1086/518226}

\bibitem[{{Camilo} {et~al.}(2007{\natexlab{c}}){Camilo}, {Ransom},
  {Pe{\~n}alver}, {Karastergiou}, {van Kerkwijk}, {Durant}, {Halpern},
  {Reynolds}, {Thum}, {Helfand}, {Zimmerman}, \& {Cognard}}]{Camilo07c}
{Camilo}, F., {Ransom}, S.~M., {Pe{\~n}alver}, J., {et~al.} 2007{\natexlab{c}},
  ApJ, 669, 561, \dodoi{10.1086/521548}

\bibitem[{{Camilo} {et~al.}(2016){Camilo}, {Ransom}, {Halpern}, {Alford},
  {Cognard}, {Reynolds}, {Johnston}, {Sarkissian}, \& {van Straten}}]{Camilo16}
{Camilo}, F., {Ransom}, S.~M., {Halpern}, J.~P., {et~al.} 2016, ApJ, 820, 110,
  \dodoi{10.3847/0004-637X/820/2/110}

\bibitem[{{Dai} {et~al.}(2019){Dai}, {Lower}, {Bailes}, {Camilo}, {Halpern},
  {Johnston}, {Kerr}, {Reynolds}, {Sarkissian}, \& {Scholz}}]{Dai19}
{Dai}, S., {Lower}, M.~E., {Bailes}, M., {et~al.} 2019, ApJ, 874, L14,
  \dodoi{10.3847/2041-8213/ab0e7a}

\bibitem[{{Duncan} \& {Thompson}(1992)}]{DT92}
{Duncan}, R.~C., \& {Thompson}, C. 1992, \apjl, 392, L9, \dodoi{10.1086/186413}

\bibitem[{{Eie} {et~al.}(2021){Eie}, {Terasawa}, {Akahori}, {Oyama}, {Hirota},
  {Yonekura}, {Enoto}, {Sekido}, {Takefuji}, {Misawa}, {Tsuchiya}, {Kisaka},
  {Aoki}, \& {Honma}}]{Eie21}
{Eie}, S., {Terasawa}, T., {Akahori}, T., {et~al.} 2021, PASJ,
  \dodoi{10.1093/pasj/psab098}

\bibitem[{{Gelfand} \& {Gaensler}(2007)}]{GG07}
{Gelfand}, J.~D., \& {Gaensler}, B.~M. 2007, \apj, 667, 1111,
  \dodoi{10.1086/520526}

\bibitem[{{Gotthelf} {et~al.}(2018){Gotthelf}, {Halpern}, {Grefenstette},
  {Harrison}, {Madsen}, \& {Miyasaka}}]{Gotthelf18}
{Gotthelf}, E.~V., {Halpern}, J.~P., {Grefenstette}, B.~W., {et~al.} 2018, The
  Astronomer's Telegram, 12297

\bibitem[{{Gupta} {et~al.}(2017){Gupta}, {Ajithkumar}, {Kale}, {Nayak},
  {Sabhapathy}, {Sureshkumar}, {Swami}, {Chengalur}, {Ghosh},
  {Ishwara-Chandra}, {Joshi}, {Kanekar}, {Lal}, \& {Roy}}]{Gupta17}
{Gupta}, Y., {Ajithkumar}, B., {Kale}, H.~S., {et~al.} 2017, Current Science,
  113, 707, \dodoi{10.18520/cs/v113/i04/707-714}

\bibitem[{{Haslam} {et~al.}(1982){Haslam}, {Salter}, {Stoffel}, \&
  {Wilson}}]{Haslam82}
{Haslam}, C.~G.~T., {Salter}, C.~J., {Stoffel}, H., \& {Wilson}, W.~E. 1982,
  A\&AS, 47, 1

\bibitem[{{Hobbs} {et~al.}(2020){Hobbs}, {Manchester}, {Dunning}, {Jameson},
  {Roberts}, {George}, {Green}, {Tuthill}, {Toomey}, {Kaczmarek}, {Mader},
  {Marquarding}, {Ahmed}, {Amy}, {Bailes}, {Beresford}, {Bhat}, {Bock},
  {Bourne}, {Bowen}, {Brothers}, {Cameron}, {Carretti}, {Carter}, {Castillo},
  {Chekkala}, {Cheng}, {Chung}, {Craig}, {Dai}, {Dawson}, {Dempsey}, {Doherty},
  {Dong}, {Edwards}, {Ergesh}, {Gao}, {Han}, {Hayman}, {Indermuehle},
  {Jeganathan}, {Johnston}, {Kanoniuk}, {Kesteven}, {Kramer}, {Leach},
  {Mcintyre}, {Moss}, {Os{\l}owski}, {Phillips}, {Pope}, {Preisig}, {Price},
  {Reeves}, {Reilly}, {Reynolds}, {Robishaw}, {Roush}, {Ruckley}, {Sadler},
  {Sarkissian}, {Severs}, {Shannon}, {Smart}, {Smith}, {Smith}, {Sobey},
  {Staveley-Smith}, {Tzioumis}, {van Straten}, {Wang}, {Wen}, \&
  {Whiting}}]{Hobbs20}
{Hobbs}, G., {Manchester}, R.~N., {Dunning}, A., {et~al.} 2020, \pasa, 37,
  e012, \dodoi{10.1017/pasa.2020.2}

\bibitem[{{Ibrahim} {et~al.}(2004){Ibrahim}, {Markwardt}, {Swank}, {Ransom},
  {Roberts}, {Kaspi}, {Woods}, {Safi-Harb}, {Balman}, {Parke}, {Kouveliotou},
  {Hurley}, \& {Cline}}]{Ibrahim04}
{Ibrahim}, A.~I., {Markwardt}, C.~B., {Swank}, J.~H., {et~al.} 2004, ApJL, 609,
  L21, \dodoi{10.1086/422636}

\bibitem[{{Joshi} {et~al.}(2018){Joshi}, {Maan}, {Surnis}, {Bagchi}, \&
  {Manoharan}}]{Joshi18}
{Joshi}, B.~C., {Maan}, Y., {Surnis}, M.~P., {Bagchi}, M., \& {Manoharan},
  P.~K. 2018, The Astronomer's Telegram, 12312

\bibitem[{{Kijak} {et~al.}(2011){Kijak}, {Lewandowski}, {Maron}, {Gupta}, \&
  {Jessner}}]{Kijak11a}
{Kijak}, J., {Lewandowski}, W., {Maron}, O., {Gupta}, Y., \& {Jessner}, A.
  2011, A\&A, 531, A16, \dodoi{10.1051/0004-6361/201014274}

\bibitem[{{Kijak} {et~al.}(2013){Kijak}, {Tarczewski}, {Lewandowski}, \&
  {Melikidze}}]{Kijak13}
{Kijak}, J., {Tarczewski}, L., {Lewandowski}, W., \& {Melikidze}, G. 2013, ApJ,
  772, 29, \dodoi{10.1088/0004-637X/772/1/29}

\bibitem[{{Lazaridis} {et~al.}(2008){Lazaridis}, {Jessner}, {Kramer},
  {Stappers}, {Lyne}, {Jordan}, {Serylak}, \& {Zensus}}]{Lazaridis08}
{Lazaridis}, K., {Jessner}, A., {Kramer}, M., {et~al.} 2008, MNRAS, 390, 839,
  \dodoi{10.1111/j.1365-2966.2008.13794.x}

\bibitem[{{Levin} {et~al.}(2012){Levin}, {Bailes}, {Bates}, {Bhat}, {Burgay},
  {Burke-Spolaor}, {D'Amico}, {Johnston}, {Keith}, {Kramer}, {Milia},
  {Possenti}, {Stappers}, \& {van Straten}}]{Levin12}
{Levin}, L., {Bailes}, M., {Bates}, S.~D., {et~al.} 2012, \mnras, 422, 2489,
  \dodoi{10.1111/j.1365-2966.2012.20807.x}

\bibitem[{{Levin} {et~al.}(2019){Levin}, {Lyne}, {Desvignes}, {Eatough},
  {Karuppusamy}, {Kramer}, {Mickaliger}, {Stappers}, \& {Weltevrede}}]{Levin19}
{Levin}, L., {Lyne}, A.~G., {Desvignes}, G., {et~al.} 2019, \mnras, 488, 5251,
  \dodoi{10.1093/mnras/stz2074}

\bibitem[{{Lewandowski} {et~al.}(2015){Lewandowski}, {Ro{\.z}ko}, {Kijak}, \&
  {Melikidze}}]{Lewandowski15b}
{Lewandowski}, W., {Ro{\.z}ko}, K., {Kijak}, J., \& {Melikidze}, G.~I. 2015,
  \apj, 808, 18, \dodoi{10.1088/0004-637X/808/1/18}

\bibitem[{{Lorimer} \& {Kramer}(2004)}]{Handbook04}
{Lorimer}, D.~R., \& {Kramer}, M. 2004, {Handbook of Pulsar Astronomy}, ed.
  R.~{Ellis}, J.~{Huchra}, S.~{Kahn}, G.~{Rieke}, \& P.~B. {Stetson}

\bibitem[{{Lyne} {et~al.}(2018){Lyne}, {Levin}, {Stappers}, {Mickaliger},
  {Desvignes}, \& {Kramer}}]{Lyne18}
{Lyne}, A., {Levin}, L., {Stappers}, B., {et~al.} 2018, The Astronomer's
  Telegram, 12284

\bibitem[{{Maan} {et~al.}(2019){Maan}, {Joshi}, {Surnis}, {Bagchi}, \&
  {Manoharan}}]{Maan19b}
{Maan}, Y., {Joshi}, B.~C., {Surnis}, M.~P., {Bagchi}, M., \& {Manoharan},
  P.~K. 2019, ApJL, 882, L9, \dodoi{10.3847/2041-8213/ab3a47}

\bibitem[{{Maan} {et~al.}(2021){Maan}, {van Leeuwen}, \& {Vohl}}]{MvLV21}
{Maan}, Y., {van Leeuwen}, J., \& {Vohl}, D. 2021, A\&A, 650, A80,
  \dodoi{10.1051/0004-6361/202040164}

\bibitem[{{Maan} {et~al.}(2013){Maan}, {Deshpande}, {Chandrashekar},
  {Chennamangalam}, {Raghavendra Rao}, {Somashekar}, {Anderson}, {Ezhilarasi},
  {Sujatha}, {Kasturi}, {Sandhya}, {Bauserman}, {Duraichelvan}, {Amiri},
  {Aswathappa}, {Barve}, {Sarabagopalan}, {Ananda}, {Beaudet}, {Bloss},
  {Dhamnekar}, {Egan}, {Ford}, {Krishnamurthy}, {Mehta}, {Minter}, {Nagaraja},
  {Narayanaswamy}, {O'Neil}, {Raja}, {Sahasrabudhe}, {Shelton}, {Srivani},
  {Venugopal}, \& {Viswanathan}}]{Maan13}
{Maan}, Y., {Deshpande}, A.~A., {Chandrashekar}, V., {et~al.} 2013, ApJS, 204,
  12, \dodoi{10.1088/0067-0049/204/1/12}

\bibitem[{{Melikidze} {et~al.}(2000){Melikidze}, {Gil}, \& {Pataraya}}]{MGP00}
{Melikidze}, G.~I., {Gil}, J.~A., \& {Pataraya}, A.~D. 2000, "ApJ", 544, 1081,
  \dodoi{10.1086/317220}

\bibitem[{{Mitra} {et~al.}(2009){Mitra}, {Gil}, \& {Melikidze}}]{MGM09}
{Mitra}, D., {Gil}, J., \& {Melikidze}, G.~I. 2009, "APJL", 696, L141,
  \dodoi{10.1088/0004-637X/696/2/L141}

\bibitem[{{Olausen} \& {Kaspi}(2014)}]{ok14}
{Olausen}, S.~A., \& {Kaspi}, V.~M. 2014, \apjs, 212, 6,
  \dodoi{10.1088/0067-0049/212/1/6}

\bibitem[{{Radhakrishnan} \& {Cooke}(1969)}]{RC69}
{Radhakrishnan}, V., \& {Cooke}, D.~J. 1969, Astrophys. Lett., 3, 225

\bibitem[{{Ransom}(2001)}]{RansomThesis}
{Ransom}, S.~M. 2001, PhD thesis, Harvard University

\bibitem[{{Ransom} {et~al.}(2002){Ransom}, {Eikenberry}, \&
  {Middleditch}}]{Ransom02}
{Ransom}, S.~M., {Eikenberry}, S.~S., \& {Middleditch}, J. 2002, AJ, 124, 1788

\bibitem[{{Remazeilles} {et~al.}(2015){Remazeilles}, {Dickinson}, {Banday},
  {Bigot-Sazy}, \& {Ghosh}}]{Remazeilles15}
{Remazeilles}, M., {Dickinson}, C., {Banday}, A.~J., {Bigot-Sazy}, M.~A., \&
  {Ghosh}, T. 2015, MNRAS, 451, 4311, \dodoi{10.1093/mnras/stv1274}

\bibitem[{{Rohlfs} \& {Wilson}(2004)}]{RW04}
{Rohlfs}, K., \& {Wilson}, T.~L. 2004, {Tools of radio astronomy}

\bibitem[{{Ruderman} \& {Sutherland}(1975)}]{RS75}
{Ruderman}, M.~A., \& {Sutherland}, P.~G. 1975, ApJ, 196, 51

\bibitem[{{Szary} {et~al.}(2015){Szary}, {Melikidze}, \& {Gil}}]{SMG15}
{Szary}, A., {Melikidze}, G.~I., \& {Gil}, J. 2015, \apj, 800, 76,
  \dodoi{10.1088/0004-637X/800/1/76}

\bibitem[{{Torne} {et~al.}(2020){Torne}, {Mac{\'\i}as-P{\'e}rez}, {Ladjelate},
  {Ritacco}, {S{\'a}nchez-Portal}, {Berta}, {Paubert}, {Calvo}, {Desvignes},
  {Karuppusamy}, {Navarro}, {John}, {S{\'a}nchez}, {Pe{\~n}alver}, {Kramer}, \&
  {Schuster}}]{Torne20}
{Torne}, P., {Mac{\'\i}as-P{\'e}rez}, J., {Ladjelate}, B., {et~al.} 2020, \aap,
  640, L2, \dodoi{10.1051/0004-6361/202038504}

\bibitem[{{Torne} {et~al.}(2022){Torne}, {Bell}, {Bintley}, {Desvignes},
  {Berry}, {Dempsey}, {Ho}, {Parsons}, {Eatough}, {Karuppusamy}, {Kramer},
  {Kramer}, {Liu}, {Paubert}, {Sanchez-Portal}, \& {Schuster}}]{Torne22}
{Torne}, P., {Bell}, G., {Bintley}, D., {et~al.} 2022, arXiv e-prints,
  arXiv:2201.07820.
\newblock \doarXiv{2201.07820}

\bibitem[{{Trushkin} {et~al.}(2019){Trushkin}, {Bursov}, {Tsybulev},
  {Nizhelskij}, \& {Erkenov}}]{Trushkin19}
{Trushkin}, S.~A., {Bursov}, N.~N., {Tsybulev}, P.~G., {Nizhelskij}, N.~A., \&
  {Erkenov}, A. 2019, The Astronomer's Telegram, 12372

\bibitem[{{van Straten} \& {Bailes}(2011)}]{vSB11}
{van Straten}, W., \& {Bailes}, M. 2011, \pasa, 28, 1, \dodoi{10.1071/AS10021}

\end{thebibliography}
\bibliographystyle{aasjournal}

\end{document}